# Building blocks of protein structures – Physics meets Biology

Tatjana Škrbić[1,2], Amos Maritan[3], Achille Giacometti[2], George D. Rose[4], Jayanth R. Banavar[1]

1. Department of Physics and Institute for Fundamental Science, University of Oregon, Eugene, OR 97403, USA
2. Dipartimento di Scienze Molecolari e Nanosistemi, Università Ca' Foscari Venezia, Campus Scientifico, Edificio Alfa, via Torino 155, 30170 Venezia Mestre, Italy
3. Dipartimento di Fisica e Astronomia, Università di Padova and INFN, via Marzolo 8, 35131 Padova, Italy
4. T. C. Jenkins Department of Biophysics, Johns Hopkins University, 3400 N. Charles Street, Baltimore, MD 21218-2683, USA

## Abstract

**The native state structures of globular proteins are stable and well-packed indicating that self-interactions are favored over protein-solvent interactions under folding conditions. We use this as a guiding principle to derive the geometry of the building blocks of protein structures – α-helices and strands assembled into β-sheets – with no adjustable parameters, no amino acid sequence information, and no chemistry. There is an almost perfect fit between the dictates of mathematics and physics and the rules of quantum chemistry. Our theory establishes an energy landscape that channels protein evolution by providing sequence-independent platforms for elaborating sequence-dependent functional**

**diversity. Our work highlights the vital role of discreteness in life and has implications for the creation of artificial life and on the nature of life elsewhere in the cosmos.**

Proteins (1-40) [we apologize that we have only included a limited selection of papers], the molecular machines of life, are formidably complex (41). They have myriad degrees of freedom, an astronomical number of possible sequences for even a moderate length chain, and are stabilized by thousands of interactions, both intra-molecular and with solvent. Yet, many proteins adopt their native conformation spontaneously under physiological conditions (5). The native state structures of globular proteins are space-filling and maximize self-interaction (6,7,9). The folded structures (21,26,32,35) are modular and built on scaffolds of α-helices (2) and strands of β-sheet (3), the only two conformers that can be extended indefinitely without steric interference while providing hydrogen-bonding partners for their own backbone polar groups (4,10,28). Proteins are digital molecules: nature's exclusion of α-β hybrid segments (27) – part α-helix, part β-strand – is built into proteins at the covalent level and restricts the topology of single domain proteins to a few thousand distinct folds at most (8,14,20).

Helices are ubiquitous in biomolecular structures. They are also found in everyday life, e.g. a garden hose (or a flexible tube) is often wound into a helix. Figure 1a is a sketch of a segment of a protein helix shown with a tube envelope. A uniform, flexible, self-avoiding solid tube, whose axis is a line, is a geometrical generalization of a sphere. A sphere is a region carving out space around a point, its center. Analogously, all points within the tube are at a distance from the tube axis smaller than or equal to the tube thickness, which is measured by the tube radius, $\Delta$. A flexible tube is an extended object with uniaxial symmetry and is not plagued by symmetry conflicts, unlike the simple model of a chain of tethered spheres for which the uniaxial symmetry inherent to a chain clashes with the spherical symmetry of the constituent objects.

Here we model a protein as a discretized tube with a set of equally spaced points, analogous to the $C_\alpha$ atoms along the protein backbone, defining its axis. The coordinates of these points are described using two angles: $\theta$ and $\mu$ (see Figure 2). The simplest repeating geometry of the axis of a tube of radius $\Delta$ is a helix of pitch P, wrapped around a straight cylinder of radius R, taken to be the helix radius. The helix is parameterized by a variable t and is defined by

$$\mathbf{r}(t) = (R\cos(t), R\sin(t), Pt/(2\pi)). \quad (1)$$

As t advances by an integer multiple of $2\pi$, the helix repeats periodically along the z-axis, with an increment equal to the pitch. The helical tube geometry is characterized by three dimensionless quantities $\Delta/R$, $\eta=P/(2\pi R)$, and $\varepsilon_0$, the rotation angle between successive points along the axis. Our initial goal parallels the seminal work of Pauling et al. (2), who sought rotation angles that allowed for the optimal placement of hydrogen bonds in a helix. The crucial difference here is that we do not need to invoke quantum chemistry, covalent bonds, the planarity of peptide bonds or hydrogen bonds.

We seek to maximize the self-interaction of a *continuum* tube (42-47) by winding the tube as tightly as possible, subject to the excluded volume constraint that the tube cannot penetrate itself. We ensure local space-filling of the helix by equating the tube radius to the local radius of curvature (Fig 1c), which, in turn, is equal to $R(1+\eta^2)$ (46) yielding:

$$\Delta = R(1+\eta^2). \quad (2)$$

The successive turns of a space-filling helix need to be parallel and alongside each other (Figure 1e). The square of the distance between a reference point in the continuum helix (denoted by $t_0$=0°) and an arbitrary point t is given by

$$d^2 = R^2 [2(1-\cos t)+\eta^2 t^2]. \quad (3)$$

We determine the parameter value $t_{min}$ for which $d^2$ is a minimum and set this minimum distance equal to the square of the tube diameter, $4\Delta^2$, thereby ensuring non-local space-filling (Figure 1f). The minimization condition is

$$\sin t_{min}+\eta^2 t_{min} = 0, \quad (4)$$

and the distance constraint is

$$4\Delta^2 = R^2[2(1-\cos t_{min})+\eta^2 t_{min}^2]. \quad (5)$$

We solve Equations (2, 4, and 5) simultaneously to obtain the unique geometry of the continuum space-filling helix (Figure 1c,e,f): $\eta$~0.4, $\Delta/R$~1.16, and $t_{min}$~302°.

The idealized continuum tube does not take into account discreteness, a common ingredient to all matter, which is crucial at small length scales. A unique benefit of discreteness is the emergence of a second building block (besides the space filling helix): a two dimensional strand with a zig-zag tube axis (Figure 3a), the rotation angle $\varepsilon_0$ of 180°, and $\mu$=180°. The existence of two building blocks is *required* for the rich diversity of topologically distinct folds, necessary for the versatile functioning of the molecular machines. A helix is defined by a repeat of ($\theta$,$\mu$)-values and a planar strand by a repeat of $\mu$=180°. For repeat $\mu$-values close to 180°, one obtains a twisted planar strand, a geometrical feature often observed in protein structures.

Figure 1g shows the space-filling discrete helix with $\eta$~0.4 and $\Delta/R$~1.16, the geometrical characteristics of the continuum space-filling helix. The discretization requires the specification of the rotation angle $\varepsilon_0$ between successive points that retains the space-filling conditions for the discrete case. This choice of $\varepsilon_0$ is made (in direct analogy with the continuum case) by requiring that the distance between points i (analogous to $t_0=0°$) and i+m with integer m (analogous to $t_{min}$) is equal to the tube diameter and the angles (i-1,i,i+m) and (i,i+m,i+m+1) are both equal to 90° (analogous to the minimization condition). The smallest value of m for which these conditions are satisfied is m=3 and $\varepsilon_0$~99.8° (the ratio of the distance to the tube diameter is found to be 1.00… and both the angles are 90.0…° for this value of $\varepsilon_0$). Upon defining the length scale to match the mean $C_\alpha$-$C_\alpha$ distance along the protein backbone of 3.81Å, the tube radius is found to be $\Delta$~2.63Å. Using these basic results, one may derive many attributes of the space-filling discrete helix, which are in excellent accord with the $\alpha$-helix building block of protein structures (see Figures 4-5, Table 1).

A space-filling helix maximizes self-interaction through local interactions, whereas the non-local interactions of strands assembled into sheets leads to space-filling. We build on the insights gained from the helix analysis to make predictions of the geometrical arrangements for strand pairing (Figure 3b-c). First, the strands need to be in phase with each other mimicking the behavior of adjoining turns in the continuum helix, placed parallel to and alongside each other. Second, there are two distinct ways (Figures 3b-c) of accomplishing space-filling of assembled strands corresponding to anti-parallel and parallel $\beta$-sheet hydrogen bonding patterns, first predicted by Pauling and Corey (3) based on hydrogen bonding. The space-filling packing requires that the distances (i,j) in Figure 3b (anti-parallel arrangement) and (i,$M_j$) in Figure 3c (parallel arrangement), which are measures of the closest approach of two parallel tube segments, both ought to be 2Δ~5.26Å (see Figure 3d-e and Table 1).

It is important to note that, for both helices and sheets, the side-chains do not clash sterically unlike in a well-packed compact arrangement of parallel strands in a hexagonal array. In addition to helices and strands, chain turns are needed to inter-connect these building blocks. In proteins, the most abundant turns are β-turns, tight, four-residue segments that approximately reverse the overall chain direction (13). β-turns are tightly wound like an α-helix, and therefore are predicted to have similar θ-angles as in the α-helix (Figure 4).

Figure 4b shows the (θ,μ) coordinates for 4 classes of residues: those that participate in α-helices, parallel β-sheets, anti-parallel β-sheets, and β-turns. The black X marks the coordinates of the predicted space-filling helix. Unsurprisingly, α-helix μ-values (49.7 ± 3.9)° are a bit lower than the theoretical prediction of 52.4° because the distance between a hydrogen-bonded donor and acceptor (N-H···O=C) can be less than their summed van der Waals radii. Of course, an ideal tube is unaffected by such chemical particulars. Nevertheless, the predicted μ value for an ideal tube is remarkably close to 50°, the average μ value for Pauling's α-helix (2), with 3.6 residues per turn. As predicted, the tight turns predominantly have a θ value close to that of the α-helix. The β strands are twisted with a μ angle around 180° and have a spread of θ angles.

The accord between our prediction and structural data from the protein data bank underscores the consilience (48) between mathematics and physics on one hand and quantum chemistry on the other and show how self-interaction is maximized through a space-filling arrangement of individual helices and sheets (Figure 6). The large but finite number of protein native state folds (8,14,20) sculpted by geometry and symmetry (24,25) is reminiscent of the restriction of the number of space groups of Bravais lattices of three-dimensional crystals to exactly 230 due to periodicity and space-filling requirements (49).

Our theory shows convincingly that structure-space and sequence-space of proteins are separable, yielding sequence-independent forms (22) that are Platonic and immutable, and not subject to Darwinian evolution. Sequences can then populate these forms resulting in the evolution of the functional diversity of life. The evolution (40,50,51) of biological macromolecules can be framed as a random walk in an inordinately vast sequence space, with selection guided by "fitness". Our formalism imposes an important constraint on protein evolution. A consequence is that the repertoire of possible folds is generated from pre-sculpted α-helices and β-strands, and, of necessity, accessible folds are mix-and-match constructs of these fundamental forms. This diversity of structural scaffolds provides a platform for elaborating functional diversity.

In seminal work, Anfinsen (5) demonstrated that proteins fold rapidly and reproducibly into their native state structures. This naturally led to the text book wisdom (35) that *the amino acid sequence of a protein determines its three-dimensional structure* leading to much effort in finding the energy minimum of a many-body complex system of a protein in its solvent with a huge number of degrees of freedom and with myriad interactions. Subsequent work by Matthews (16) and others showed that protein structure is nevertheless *very tolerant of amino acid replacement.*

Our results here conclusively demonstrate a simple two-step process for understanding proteins. First, a menu of putative native state structures is created without regard to amino acid sequence and chemistry. In the second step, a given protein selects its native state from this menu. Thus the horrendous problem of working out the native state structure of a given protein from knowledge of its sequence by finding, from scratch, the conformation, which minimizes the net energy of myriad imperfectly known microscopic interactions, is replaced by the much simpler task of finding the best fit of the sequence to one among the library of geometrically sculpted folds determined in a sequence-independent and chemistry-independent manner. This best-fit process, also exploited in the threading algorithm (15), is where the role of the amino acid sequence becomes paramount. Indeed, in an influential series of papers (12,17-19), it has been highlighted that the amino acid side chains must be able to fit into the native state fold with minimal frustration thereby creating a landscape akin to a folding funnel.

Some 80 years ago, Bernal (1) wrote – *Any effective picture of protein structure must provide at the same time for the common character of all proteins as exemplified by their many chemical and physical similarities, and for the highly specific nature of each protein type. It is reasonable to believe, though impossible to prove, that the first of these depends on some common arrangement of the amino acids*. Indeed, our work here shows that the common character of all proteins originates from an appropriate tube-like geometrical description of just the backbone $C_\alpha$ atoms, which are common to all proteins, and results in the library of native state folds sculpted by geometry and symmetry, without a need for sequence specificity or chemistry. *The highly specific nature of each protein type* then arises from its distinctive amino acid side-chains and their fit to one of the folds from the library. For a protein, the folded structure is central to its functionality. The situation is loosely analogous to a restaurant in which the chef (geometry and symmetry) creates a menu of items (the library of putative native state folds) that customers (protein sequences) can order from (fold into). The chef does not cater to the individual tastes of the customers. Rather, all patrons of the restaurant are satisfied picking an item from the menu. As in proteins, the total number of patrons can vastly exceed the number of menu items. If, in fact, the menu of protein structures itself evolved, then one would be confronted by an almost impossible situation for evolution and natural selection in which a protein and its interacting partners would have to co-evolve their structures synergistically in order to maintain function. This situation is deftly avoided by the geometrically determined native state folds providing a fixed backdrop for evolution to shape protein sequences and functionalities.

Richard Feynman, in a lecture entitled *There's Plenty of Room in the Bottom: An Invitation to Enter a New Field of Physics* at the annual American Physical Society Meeting at Caltech on December 29, 1959, suggested that tiny, nanoscale machines could be constructed by manipulating individual atoms. Proteins are precisely such machines (21,26,32,35). Indeed, proteins as well as macroscopic machines establish a stable framework that can accommodate moving parts, which perform a function. Proteins are nature's implementation of the abstract forms presented here, a diversity of stable forms deduced entirely from mathematical considerations. These predictions – independent of any chemistry – have implications for life elsewhere in our cosmos (52) suggesting that there is no absolute need for carbon chemistry for life to exist. We look forward to other implementations in the lab, raising the prospect of powerful interacting machines, potentially leading to artificial life (53).

In summary, underlying life's evolving complexity (41) is a sequence-independent energy landscape with thousands of stable minima — a landscape formed from nature's scaffold building blocks, a protein grammar. In both natural and artificial languages, a grammar is a finite set of rules that can generate an a large number of syntactically correct sentences or strings. The discretized tube model establishes an immutable grammar of life and "*from so simple a beginning, endless*" – protein sequences and functionalities – *"most beautiful and most wonderful have been, and are being, evolved"* (54).

**PDB analysis:** We have carried out a quantitative comparison between our predictions and protein structure. To develop a working set for comparison, Richardsons' Top 8000 set of high-resolution, quality-filtered protein chains (resolution < 2Å, 70% PDB homology level) [see the web site: http://kinemage.biochem.duke.edu/databases/top8000.php ] was further filtered to exclude all structures with missing backbone atoms, yielding a working set of 4416 structures (listed in Table 2). The working set was cross-checked against 478 proteins having a more stringent homology cutoff of 20%, taken from the Pisces database (23); 205 entries are in common to both sets. Almost all bond lengths ($C_{\alpha(i)}$-$C_{\alpha(i+1)}$ distance) (~99.7%) in the working set are clustered around 3.81Å, as expected for a *trans* peptide. Those remaining have shorter bonds, ~2.95Å, predominantly from *cis* residues. For purposes of comparison, a fixed bond length of 3.81Å is used. Hydrogen bonds were identified using DSSP (11). Hydrogen-bonded conformers extracted from the working set include 3595 helices, 8473 antiparallel pairs, 4639 parallel pairs, and 58,820 turns. Helices were identified as 12-residue segments with intra-helical hydrogen bonds ($N_i$-H•••$O_{i-4}$ and $O_i$•••H-$N_{i+4}$) at each residue. Antiparallel strand pairs were identified by three inter-pair hydrogen bonds at (i,j), (i+2,j-2), and (i-2,j+2), i ∊ strand1, j ∊ strand2. To avoid possible end effects, only (i,j) residue pairs were used. Parallel strand pairs were identified by four inter-pair hydrogen bonds between (i,j-1), (i,j+1), (i+2,j+1), and (i-2,j-1), $i \in$ strand1, $j \in$ strand2, and again only the i-th residue was retained. Double-counting was assiduously avoided. β turns were identified by hydrogen bonds between (i,i+3) with no helical residues among the 4. The (θ,μ)-values were then recorded for points i+1 and i+2 in the turns.

**Acknowledgements:** We are indebted to Pete von Hippel for his warm hospitality and to him, Jeremy Berg, and Brian Matthews for stimulating comments. **Funding:** This project received funding from the European Union's Horizon 2020 research and innovation program under the Marie Skłodowska-Curie Grant Agreement No 894784. The contents reflect only the authors' view and not the views of the European Commission. Support from the University of Oregon (through a Knight Chair to JRB), NSF (GDR), University of Padova through "Excellence Project 2018" of the Cariparo foundation (AM), MIUR PRIN-COFIN2017 *Soft Adaptive Networks* grant 2017Z55KCW and COST action CA17139 (AG) is gratefully acknowledged. The computer calculations were performed on the Talapas cluster at the University of Oregon.

**Author contributions:** JRB, AM and TŠ developed the ideas for the calculations, and all the authors participated in exploring the consequences of the theory. TŠ carried out the calculations under the guidance of JRB. JRB and GDR wrote the paper. All authors reviewed the manuscript.

Emails: tskrbic@uoregon.edu, amos.maritan@pd.infn.it, achille.giacometti@unive.it, grose@jhu.edu, banavar@uoregon.edu

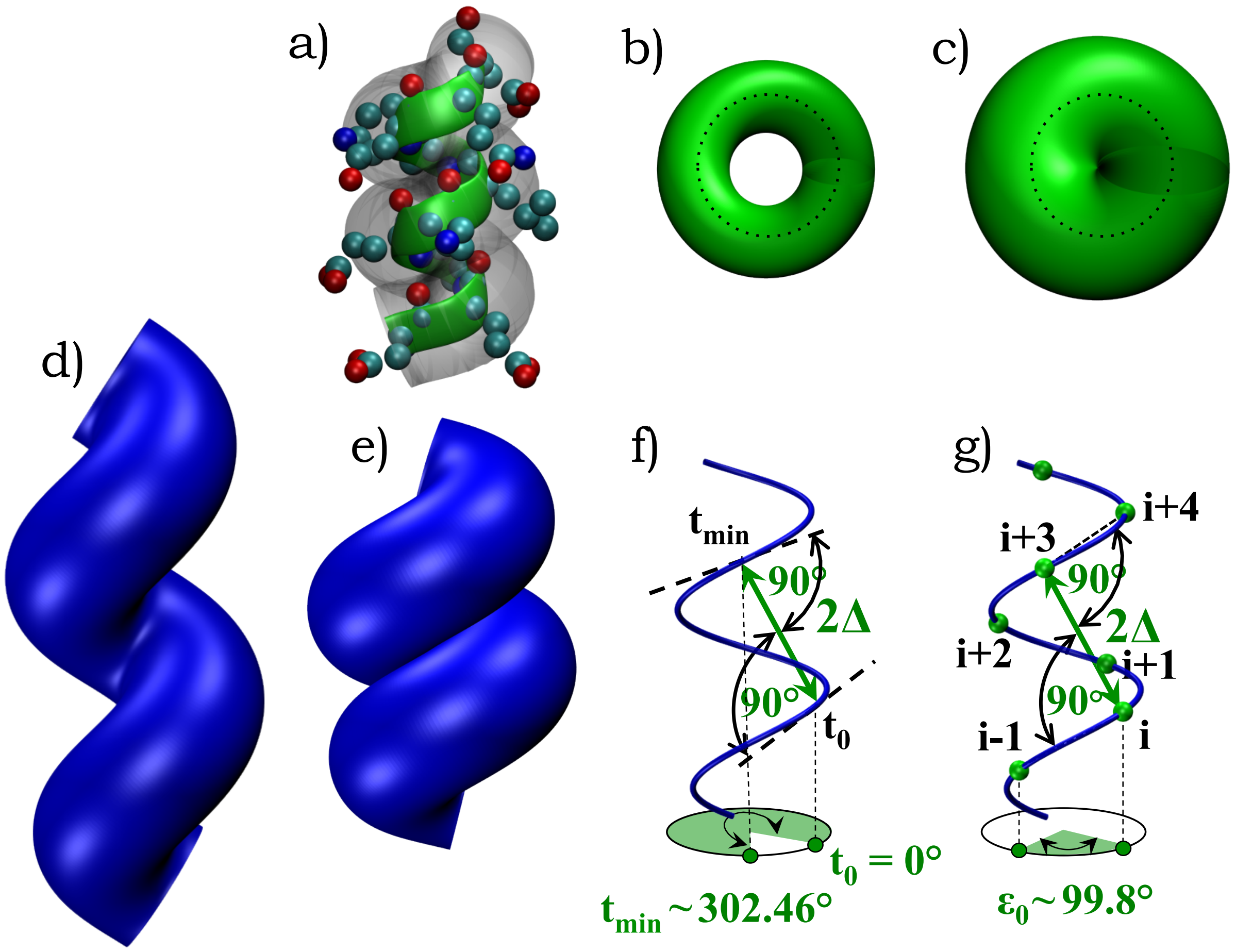

**Figure 1: Optimal geometry of space-filling helix.** (a) A segment of ten residues of a helix from phage T4 lysozyme protein 1L56 (residues 61-70). The green ribbon represents the helical trace formed by the $C_\alpha$ atoms, the spheres denote the heavy backbone and side-chains atoms in the helix, and the transparent tube is a guide to the eye. (b-c) show top-views of two continuum helices, both with a helix pitch P to helix radius R ratio $\eta=(P/2\pi R)\sim0.4$ and a local radius of curvature of the helix, $R_{local}=R(1+\eta^2)\sim1.16R$. The tube radii Δ in the two cases are different: $\Delta/R_{local}=1/2$ and 1 respectively. (b) When Δ is less than $R_{local}$, there is empty space in the interior. When Δ is bigger than $R_{local}$, the turn is too tight leading to a kink, as is sometimes observed in a garden hose (not shown). (c) The sweet spot occurs when $\Delta=R_{local}$, leading to maximization of the *local* self-interaction. (d-e) shows side views of two helices with η-values of 0.8 and ~0.4

respectively. In both cases, $\Delta$ has been chosen to be the local radius of curvature of the latter helix ~1.16R. (d) When $\eta$ is larger than ~0.4, there is empty space between successive turns and the *non-local* self-interaction is not maximized. In the other limit of small $\eta$ (not shown), successive turns of the tube overlap and this is forbidden sterically. (e) A Goldilocks situation here is when $\eta$ is tuned just right to ~0.4 yielding ($\Delta$/R)~1.16 for a continuum space-filling helix maximizing both local and non-local self-interaction. The top and side views of the optimal continuum helix are shown in (c) and (e) respectively. (f) and (g) show how these results can be captured analytically (see text) for a continuum and a discrete tube respectively.

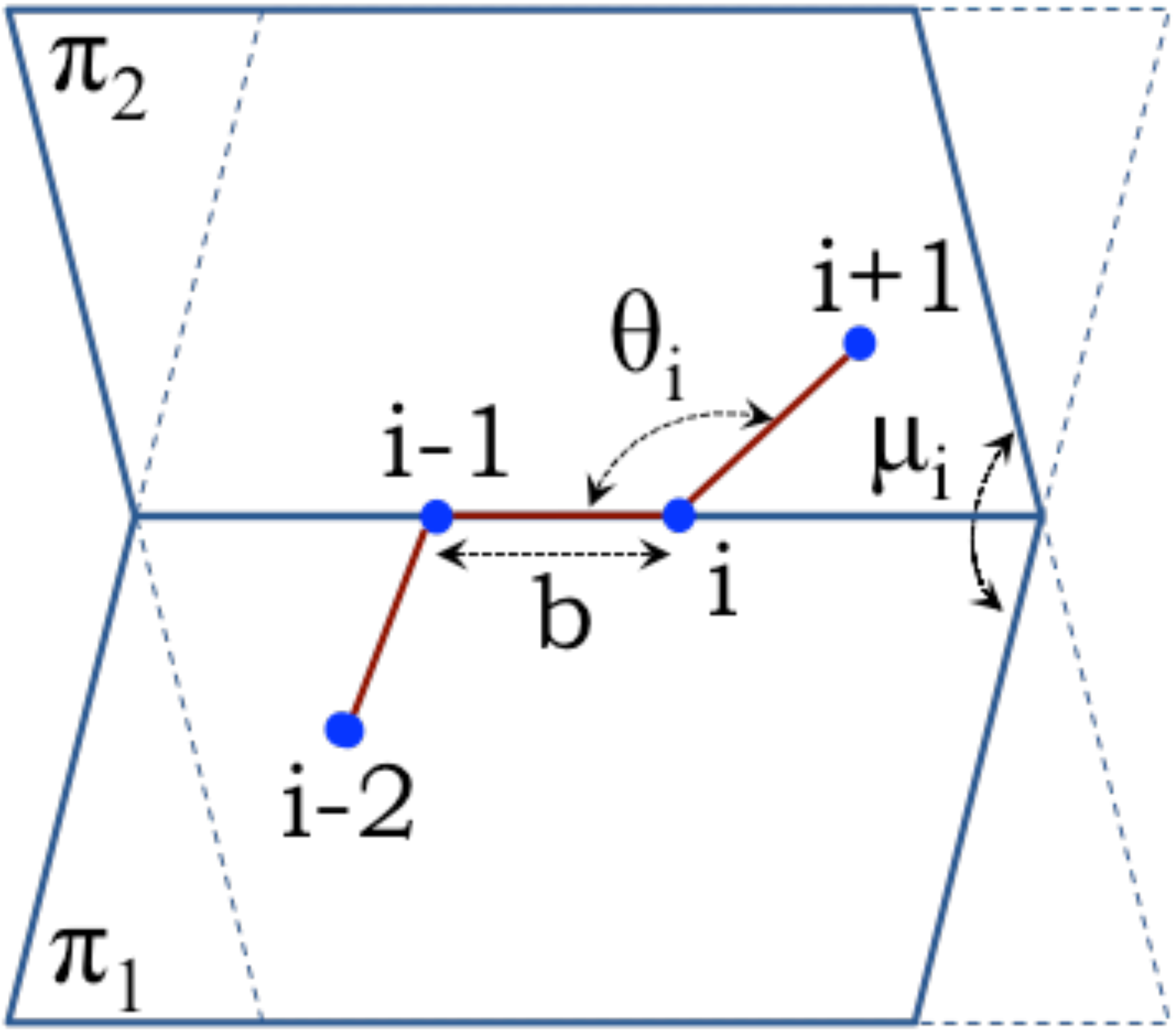

**Figure 2: Coordinate system at discrete location i along tube axis.** The bond length b, assumed here to be a constant, is the distance between successive points. The angle $\theta_i$ is the angle subtended at i by points (i-1) and (i+1) along the tube axis. $\mu_i$ is the dihedral angle between the planes $\pi_1$ and $\pi_2$ formed by [(i-2),(i-1),i] and [(i-1),i,(i+1)] respectively or equivalently the angle between the binormals in a Frenet reference frame at points (i-1) and i. Knowledge of the coordinates of the previous three points (i-2,i-1,i) and the variables ($\theta_i$, $\mu_i$) are sufficient to uniquely specify the coordinates of the point (i+1).

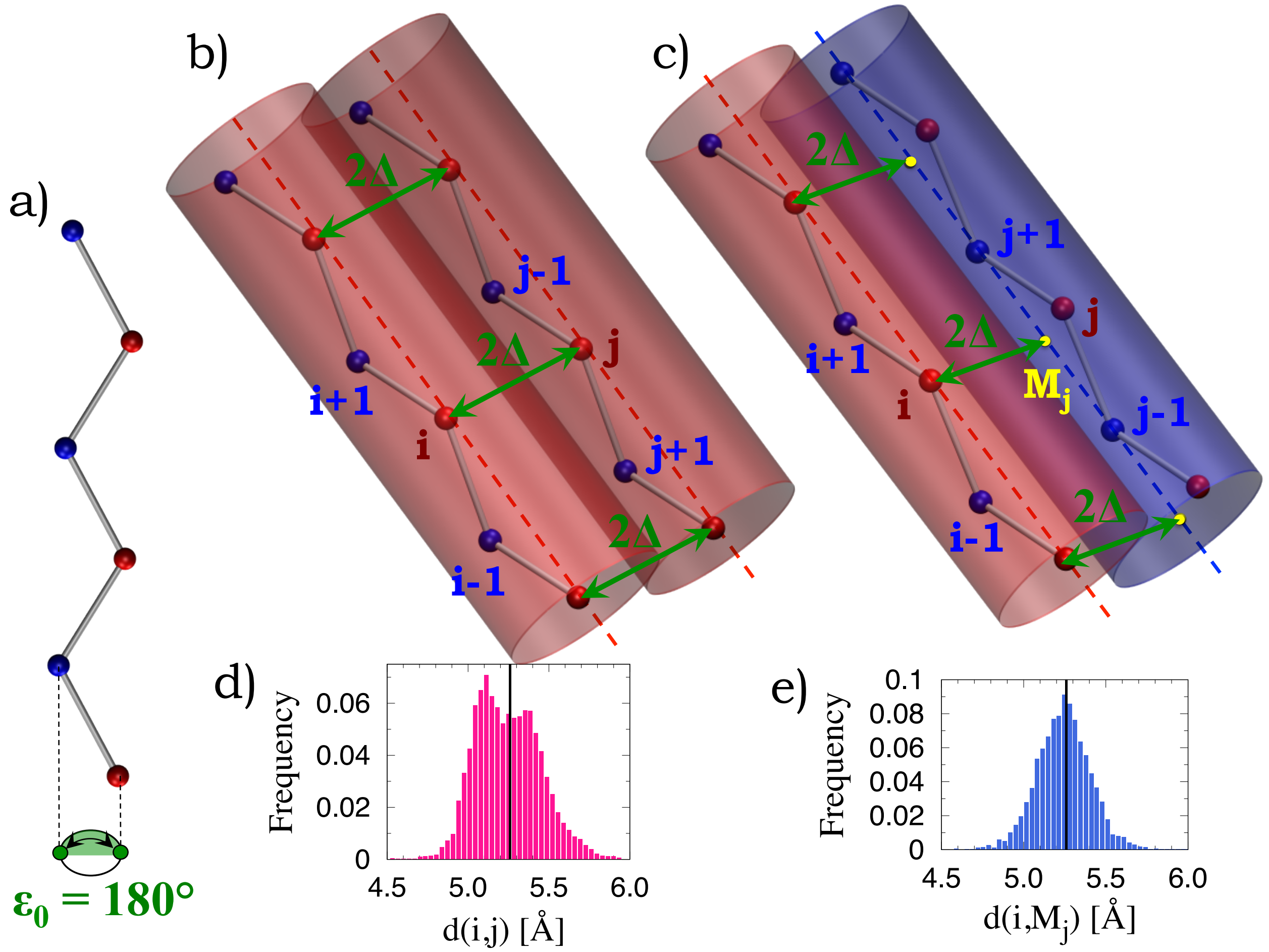

**Figure 3: Optimal packing of strands.** (a) A single two dimensional zig-zag strand (with a rotation angle of 180°) lying in the plane of the paper. This planarity can only occur for a discrete tube and is forbidden for a tube in the continuum. Alternate points along a strand are colored red and blue. There are two equivalent choices for a straight tube axis, one lying along the line of blue points (the blue axis) or the line of red points (red axis). Two distinct space-filling arrangements for strand packing are shown corresponding to (b) red axis-red axis (or equivalently blue axis-blue-axis – not shown) packing and (c) red axis-blue axis (or equivalently blue axis-red axis – not shown) packing. The two cases correspond to anti-parallel and parallel β-sheets with distinct distance constraints. The yellow point $M_j$ lies midway between the blue points j-1 and j+1. The maximization of self-interaction dictates that the distances (i,j) in (b) and (i,$M_j$) in (c) ought to be 2Δ~5.26Å to ensure space filling. (d) and (e) show the histograms of the distances (i,j) and (i,$M_j$) in the interior of anti-parallel and parallel β-sheets in protein structures. The black vertical lines show the theoretical prediction of 2Δ~5.26Å. The mean values of both histograms are the same as the theoretical prediction (see Table 1).

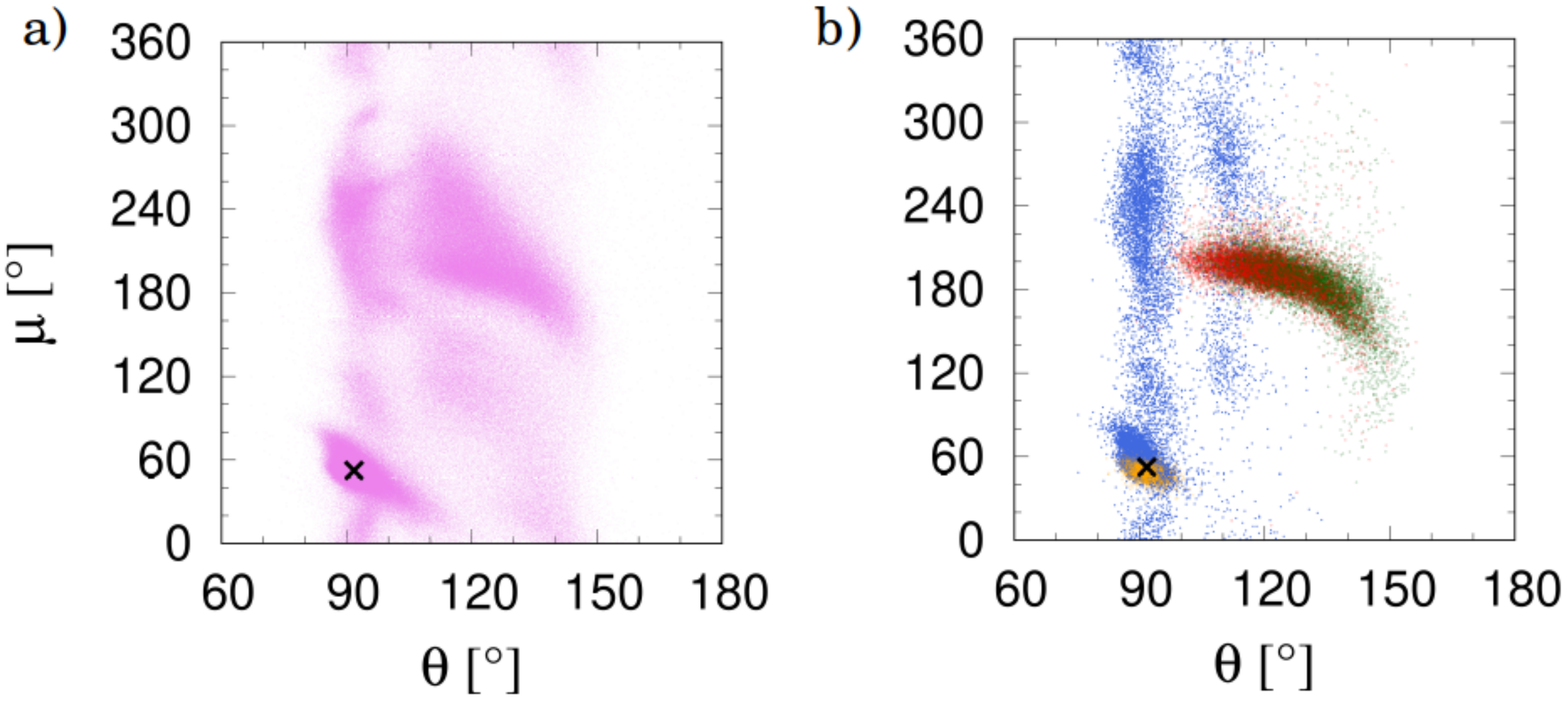

**Figure 4: Two views of the local structure representation of proteins.** a) (θ, μ) plot of the PDB data set (see Table 2) comprising 4416 proteins and 972,519 residues. Here, the local conformations of residues are shown in the (θ,μ) plane. For strands, a μ-value that deviates from ~180° is the signature of a twisted strand, which is still locally planar. The plot shows chiral symmetry breaking, i.e., the points are not symmetrically placed around μ=180°. Our simplified analysis does not attempt to account for this. b) (θ, μ) coordinates of random samples of 12000 points each from the interior of α-helices (orange); anti-parallel (green) and parallel (red) β-sheets; and β-turns (the two interior sites of (i,i+3) hydrogen-bonded residues with no helical residues) (blue). The tight turns have θ-values similar to those of helices. Unlike for helices and turns, the θ-values of strands are not constrained. The black X in both panels shows our prediction of the geometry of space-filling helices.

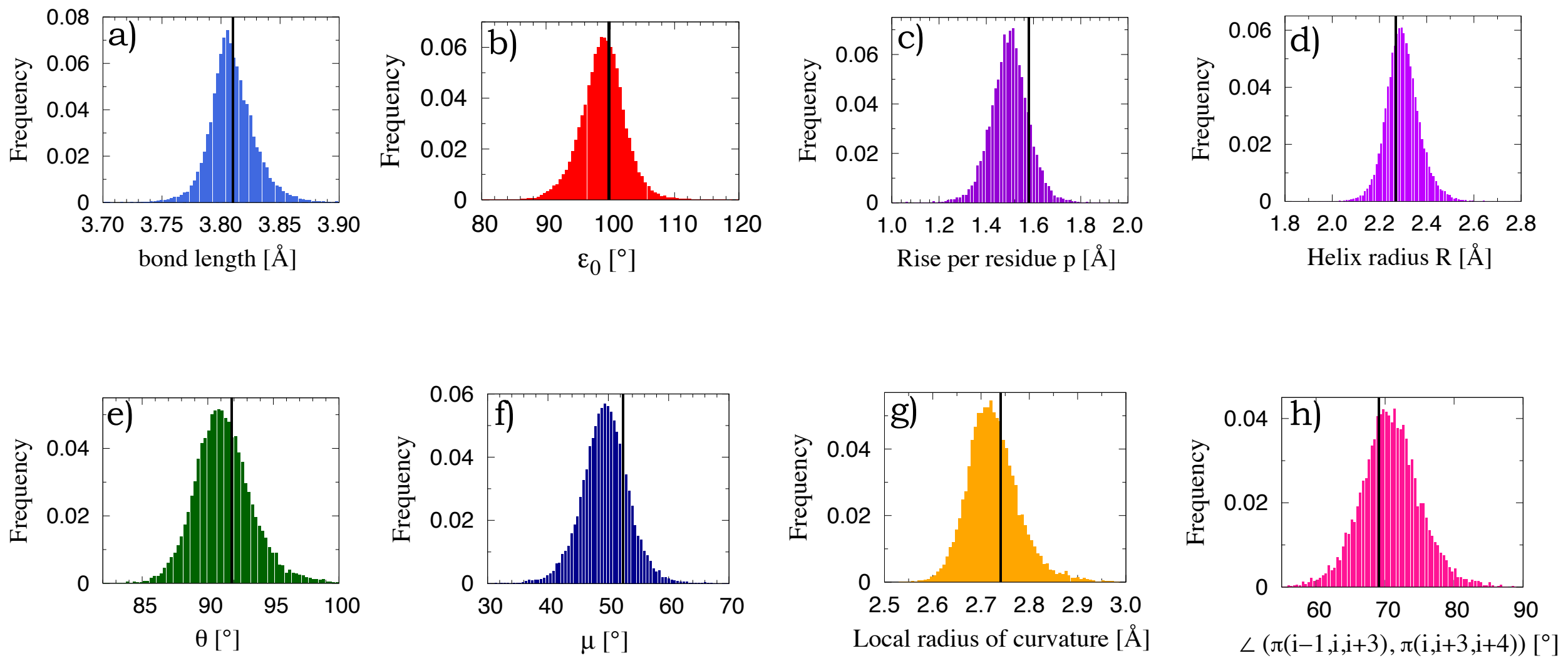

**Figure 5: Distribution of α-helix characteristics.** (a) Distribution of the experimentally determined bond lengths (consecutive $C_\alpha$-$C_\alpha$ distances). The bond length in the theory was chosen to be the mean bond length of 3.81Å and sets the characteristic length scale. The other panels show the distributions of (b) the rotation angle, (c) the rise per residue, (d) the helix radius, (e) θ, (f) μ, (g) the local radius of curvature, and (h) the dihedral angle between the planes defined by the points (i-1,i,i+3) and (i,i+3,i+4) in Figure 1g. The triangles formed by the two triplets ought to be congruent but they are not co-planar. The black line in each of the panels (except the first) shows the zero parameter theoretical prediction. Overall, there is excellent accord between theory and observations from protein structures.

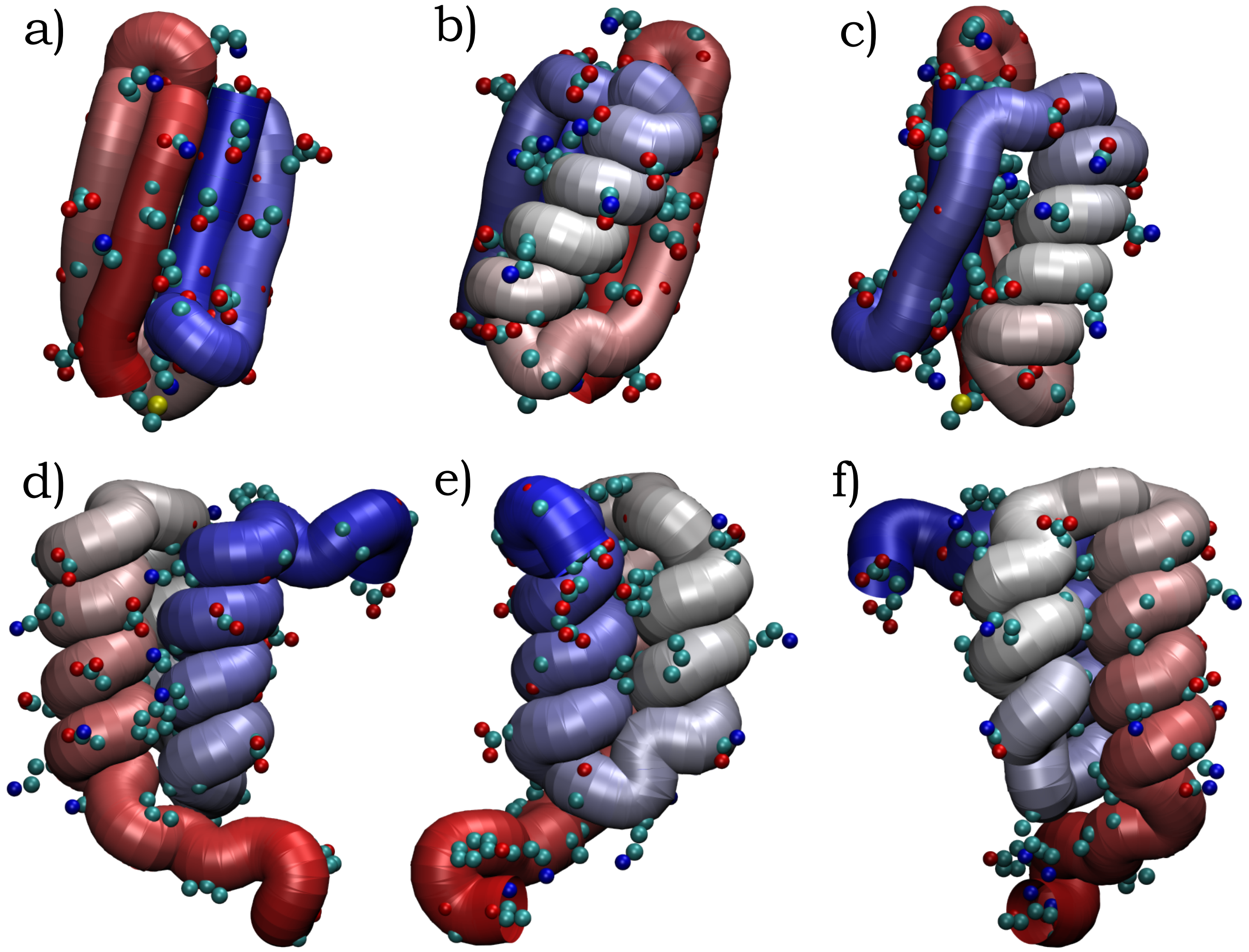

**Figure 6: Consilience between mathematics and biochemistry.** The figure shows three views each of two short proteins. (a-c) is the 56-residue long protein 3GB1 comprising 4 strands assembled into sheets along with a single helix. (d-f) is a protein of the same length, 2KDL, comprised of a three-helix bundle. Each panel shows a uniform tube, with the theoretically predicted radius of 2.6Å, whose axis passes through the $C_\alpha$ atoms. The sole exception is the β-sheet (for which hydrogen bonding was identified using DSSP (11)), where every other $C_\alpha$ atom is considered (as explained in Figures 3b and c). The tube color varies continuously from red to blue (via grey) as its axis moves from the N-terminal to the C-terminal. The heavy atoms of the side chains sticking outside the tube are shown. The maximization of the self-interaction through space-filling is evident.

| Continuum tube diameter from theory 2Δ=5.26… Å | | |
|---|---|---|
| **Quantity** | **Theory** | **PDB data** |
| **HELIX** | | |
| Rotation angle ε [°] | 99.8 | 99.1 ± 3.4 |
| Number of residues per turn | 3.61 | 3.63 ± 0.13 |
| Helix radius R [Å] | 2.27 | 2.30 ± 0.07 |
| Rise per residue p [Å] | 1.58 | 1.51 ± 0.08 |
| Helix pitch P [Å] | 5.69 | 5.47 ± 0.49 |
| Pitch to radius ratio $\eta = P/(2\pi R)$ | 0.400 | 0.377 ± 0.046 |
| ∠ (π(i-1,i,i+3), π(i,i+3,i+4)) [°] | 69.1 | 70.0 ± 4.4 |
| Local radius of curvature [Å] | 2.74 | 2.73 ± 0.05 |
| θ [°] | 91.8 | 91.3 ± 2.2 |
| μ [°] | 52.4 | 49.7 ± 3.9 |
| **SHEET** | | |
| **Type I β-sheet: parallel** | | |
| θ [°] | flexible | 121 ± 10 |
| μ [°] | ~180 | 191 ± 17 |
| d (i,$M_j$) [Å] | 2Δ=5.26 | 5.26 ± 0.16 |
| **Type II β-sheet: antiparallel** | | |
| θ [°] | flexible | 127 ± 10 |
| μ [°] | ~180 | 186 ± 20 |
| d (i,j) [Å] | 2Δ=5.26 | 5.26 ± 0.20 |

**Table 1: Quantitative comparison between theory and data from the Protein Data Bank (PDB).** We choose the bond length to match the experimentally determined mean distance between successive $C_\alpha$ atoms of 3.81 ± 0.02Å. The chain is defined by discrete points denoted by 1,2,3,…,i,… d(i,j) is the distance between the points i and j. The angle ∠(π(i,j,k), π(l,m,n)) is the dihedral angle between the two planes formed by the sites (i,j,k) and (l,m,n). $M_j$ is defined to be the geometrical center of the points j-1 and j+1. The agreement between theory and data is striking considering that the theory is parameter-free.

**Table 2: PDB codes of the 4416 proteins used in our analysis.**

| | | | | | | | | | |
|---|---|---|---|---|---|---|---|---|---|
| 16pk_A | 1iqc_C | 1pnc_A | 1w0p_A | 2buw_B | 2hwn_D | 2rc3_A | 2zk9_X | 3euf_D | 3kl0_B |
| 1a1i_A | 1iqq_A | 1pnd_A | 1w0u_A | 2bv2_B | 2hxm_A | 2rc8_B | 2zkd_B | 3eul_A | 3kl6_B |
| 1a2p_B | 1iqz_A | 1pp0_C | 1w1h_C | 2bv4_A | 2hxp_A | 2rci_A | 2zl6_B | 3eun_A | 3klq_A |
| 1a2y_A | 1irq_A | 1psr_A | 1w2c_A | 2bw0_A | 2hxs_A | 2rcq_A | 2znd_A | 3eup_B | 3klr_A |
| 1a2y_B | 1isp_A | 1ptq_A | 1w2i_B | 2bw8_A | 2hxt_A | 2rcv_E | 2znr_A | 3evf_A | 3kmt_C |
| 1a2z_C | 1isu_A | 1puc_A | 1w3i_A | 2bwf_A | 2hy5_A | 2rcz_B | 2zoo_A | 3evk_D | 3kmv_D |
| 1a34_A | 1it2_B | 1puf_B | 1w3w_A | 2bwl_A | 2hy5_B | 2rdh_C | 2zpd_A | 3evy_B | 3knb_B |
| 1a3a_A | 1itw_D | 1pvm_A | 1w3y_A | 2bwr_B | 2hy7_A | 2rdq_A | 2zpo_A | 3ew0_A | 3knv_A |
| 1a4i_B | 1itx_A | 1pvx_A | 1w4s_A | 2c0c_A | 2hyk_A | 2rdu_A | 2zpu_A | 3ew1_D | 3kp8_A |
| 1a73_A | 1iu8_B | 1pxv_B | 1w4t_A | 2c0h_A | 2hyv_A | 2rdz_A | 2zqe_A | 3ewi_A | 3kpb_D |
| 1a7d_A | 1iue_B | 1pyo_B | 1w4v_B | 2c0r_B | 2hzl_B | 2ree_A | 2zqm_A | 3exe_D | 3kq0_A |
| 1a7t_B | 1iuz_A | 1pzs_A | 1w4x_A | 2c0z_A | 2hzy_B | 2reg_A | 2zqn_B | 3exr_A | 3kqi_A |
| 1a88_A | 1iv3_D | 1q08_B | 1w53_A | 2c1d_D | 2i0q_A | 2rem_B | 2zs0_A | 3ey6_A | 3kqr_A |
| 1a8q_A | 1iv9_A | 1q0q_A | 1w5r_B | 2c1s_A | 2i1n_A | 2rer_A | 2zs0_D | 3eye_A | 3kre_A |
| 1a8s_A | 1iwd_A | 1q0r_A | 1w66_A | 2c1v_B | 2i24_N | 2rfg_A | 2zs1_B | 3eyi_A | 3krs_A |
| 1a92_C | 1ix1_B | 1q1r_B | 1w6s_C | 2c29_F | 2i2q_A | 2rfm_B | 2zs1_C | 3eyp_B | 3kru_A |
| 1ab1_A | 1ixg_A | 1q1u_A | 1w6s_D | 2c2n_A | 2i3f_A | 2rh2_A | 2zsi_A | 3ezi_B | 3kse_D |
| 1aba_A | 1iy8_C | 1q2h_A | 1w70_A | 2c2p_A | 2i49_A | 2rh3_A | 2ztl_C | 3f0y_C | 3ksh_A |
| 1afb_3 | 1iyb_A | 1q4u_B | 1w8o_A | 2c2u_A | 2i4a_A | 2rhi_A | 2zu1_B | 3f17_A | 3ksv_A |
| 1ag9_B | 1iye_C | 1q5m_B | 1w8u_A | 2c3n_C | 2i5r_B | 2rhk_C | 2zu2_A | 3f1l_A | 3ksx_A |
| 1agy_A | 1iyn_A | 1q5z_A | 1w99_A | 2c41_F | 2i5v_O | 2ri0_B | 2zux_B | 3f1p_A | 3kt9_A |
| 1ah7_A | 1izc_A | 1q6o_A | 1w9p_A | 2c42_B | 2i61_A | 2ri7_A | 2zuy_A | 3f1p_B | 3ktz_A |
| 1aho_A | 1ize_A | 1q7l_A | 1w9s_A | 2c4e_A | 2i62_D | 2ri9_A | 2zw2_A | 3f2e_A | 3ku3_B |
| 1aii_A | 1j05_B | 1q7l_B | 1wa3_A | 2c4f_T | 2i6v_A | 2rik_A | 2zwd_A | 3f2u_A | 3kus_B |
| 1ako_A | 1j0h_B | 1q8f_A | 1wb0_A | 2c4j_D | 2i7c_C | 2riq_A | 2zwj_A | 3f3q_A | 3kuv_A |
| 1aky_A | 1j0p_A | 1qau_A | 1wb6_B | 2c4n_A | 2i7d_A | 2rji_A | 2zwn_A | 3f3x_A | 3kwe_A |
| 1aoh_B | 1j1y_A | 1qav_A | 1wba_A | 2c53_A | 2i7f_B | 2rjw_A | 2zwu_A | 3f47_A | 3kxt_A |
| 1aoz_A | 1j24_A | 1qaz_A | 1wbe_A | 2c6q_B | 2i8t_B | 2rk3_A | 2zx2_A | 3f4m_A | 3kyj_A |
| 1arb_A | 1j27_A | 1qb5_E | 1wbh_B | 2c6u_A | 2i9a_D | 2rk5_A | 2zxj_B | 3f4s_A | 3kz5_A |
| 1ast_A | 1j2j_B | 1qb7_A | 1wbi_H | 2c6z_A | 2i9i_A | 2rkl_A | 2zxy_A | 3f52_A | 3kz7_A |
| 1atg_A | 1j2r_A | 1qba_A | 1wbj_A | 2c78_A | 2iax_A | 2rkq_A | 2zya_B | 3f5l_B | 3kzj_A |
| 1atl_B | 1j30_B | 1qcx_A | 1wbj_B | 2c7p_A | 2ib8_A | 2rku_A | 2zyh_B | 3f5o_G | 3kzu_B |
| 1atz_B | 1j34_A | 1qd1_B | 1wc2_A | 2c81_A | 2ibj_A | 2rky_C | 2zyo_A | 3f6o_A | 3l07_B |
| 1aun_A | 1j34_B | 1qd2_A | 1wc9_A | 2c82_B | 2ibl_A | 2sak_A | 2zzd_E | 3f6q_A | 3l0f_A |
| 1avb_A | 1j3w_C | 1qd9_C | 1wcf_A | 2c8h_D | 2ibp_B | 2sec_I | 2zzd_J | 3f6q_B | 3l0l_B |
| 1awd_A | 1j48_A | 1qdd_A | 1wcg_B | 2c92_D | 2ic6_A | 2sga_A | 2zzj_A | 3f6y_A | 3l18_A |
| 1aye_A | 1j71_A | 1qfv_B | 1wck_A | 2c95_B | 2ic7_B | 2sn3_A | 2zzr_A | 3f74_B | 3l1e_A |
| 1b0b_A | 1j75_A | 1qgi_A | 1wd3_A | 2c9q_A | 2idl_B | 2tnf_B | 2zzs_O | 3f75_A | 3l2c_A |
| 1b16_A | 1j77_A | 1qgj_A | 1wdd_S | 2cal_A | 2if6_A | 2uuy_B | 2zzv_B | 3f75_P | 3l32_A |
| 1b1c_A | 1j7d_A | 1qgu_D | 1wdy_A | 2car_B | 2ifc_C | 2uv4_A | 3a02_A | 3f7l_A | 3l39_A |
| 1b2s_F | 1j7g_A | 1qh5_B | 1wf3_A | 2cb5_A | 2ig8_A | 2uvj_A | 3a03_A | 3f7q_A | 3l3u_A |
| 1b37_B | 1j8e_A | 1qhf_A | 1whi_A | 2cb8_A | 2igi_A | 2uvo_B | 3a04_A | 3f8m_B | 3l41_A |
| 1b3a_B | 1j8u_A | 1qho_A | 1wka_A | 2cbz_A | 2igp_A | 2uw1_A | 3a07_A | 3f97_A | 3l42_A |
| 1b4f_B | 1j9l_A | 1qhq_A | 1wko_A | 2cc6_A | 2igv_A | 2uwa_A | 3a09_A | 3f9b_A | 3l46_A |
| 1b5e_A | 1ja9_A | 1qhv_A | 1wkq_B | 2cch_B | 2igx_A | 2uyt_A | 3a0y_B | 3f9r_A | 3l4p_A |
| 1b63_A | 1jae_A | 1qj5_B | 1wkr_A | 2ccq_A | 2ih5_A | 2uyw_A | 3a16_C | 3fas_B | 3l4r_A |
| 1b66_A | 1jak_A | 1qjc_B | 1wku_B | 2ccw_A | 2ihd_A | 2uyz_A | 3a1c_A | 3fb9_A | 3l5l_A |
| 1b67_A | 1jat_A | 1qjw_B | 1wkx_A | 2cdn_A | 2ii2_A | 2uyz_B | 3a21_A | 3fbg_A | 3l6g_A |
| 1b8a_B | 1jay_B | 1qkk_A | 1wld_A | 2cf7_C | 2iid_A | 2uz1_D | 3a2q_A | 3fbl_A | 3l6n_A |
| 1b8d_K | 1jcd_A | 1ql0_B | 1wlg_B | 2cfe_A | 2ijh_A | 2uzc_C | 3a2v_I | 3fd7_B | 3l77_A |
| 1b8p_A | 1jcv_A | 1ql3_B | 1wlz_C | 2cg7_A | 2ijq_A | 2v09_A | 3a2z_A | 3fde_B | 3l7h_B |
| 1b93_A | 1jd0_B | 1qlw_A | 1wm2_A | 2cgq_A | 2ijx_D | 2v0h_A | 3a39_A | 3fdl_A | 3l7t_B |
| 1bas_A | 1jd1_C | 1qmy_C | 1wma_A | 2chc_B | 2imf_A | 2v0s_A | 3a3d_B | 3fdq_A | 3l8e_B |
| 1baz_A | 1jd5_A | 1qnj_A | 1wmd_A | 2cia_A | 2imi_B | 2v0u_A | 3a3v_A | 3fdr_A | 3l8w_A |
| 1bdo_A | 1jdh_B | 1qnn_C | 1wmh_A | 2ciu_A | 2imq_X | 2v1o_B | 3a40_X | 3fe0_A | 3l91_A |
| 1beh_A | 1jdl_A | 1qnp_A | 1wmw_A | 2ciw_A | 2in8_A | 2v1q_A | 3a4r_A | 3fe7_A | 3l91_B |
| 1bf6_A | 1jek_A | 1qnx_A | 1wmz_D | 2cj3_A | 2inc_A | 2v1w_B | 3a4u_A | 3fev_A | 3l9a_X |
| 1bgf_A | 1jev_A | 1qoz_B | 1wn2_A | 2cj4_A | 2inc_B | 2v25_A | 3a4w_B | 3ff5_B | 3l9f_D |
| 1bgp_A | 1jf8_A | 1qre_A | 1wny_A | 2cjj_A | 2ior_A | 2v27_A | 3a57_A | 3ff7_C | 3l9s_A |
| 1bhp_A | 1jfl_B | 1qrp_E | 1wo8_D | 2cjl_B | 2ioy_B | 2v2g_C | 3a5p_D | 3ff9_B | 3l9u_A |
| 1bj7_A | 1jfr_A | 1qs1_A | 1wod_A | 2cjp_A | 2ip2_B | 2v33_B | 3a5r_A | 3fg0_F | 3l9y_B |
| 1bkp_A | 1jfu_A | 1qsa_A | 1wog_E | 2cjs_C | 2ip6_A | 2v36_D | 3a6r_B | 3fgd_A | 3las_B |
| 1bn8_A | 1jfx_A | 1qsg_A | 1woq_B | 2ckf_D | 2ipr_B | 2v3g_A | 3a72_A | 3fh2_A | 3lat_A |
| 1bq8_A | 1jg1_A | 1qt9_A | 1wor_A | 2ckk_A | 2iq7_A | 2v3s_A | 3a7l_A | 3fhg_A | 3lbe_D |
| 1bqb_A | 1jhd_A | 1qtn_A | 1wpa_A | 2cks_A | 2iqj_A | 2v4n_A | 3a7n_A | 3fid_A | 3lbf_C |
| 1bqk_A | 1jhf_A | 1qtw_A | 1wpn_B | 2cm4_A | 2iru_B | 2v4v_A | 3a8g_B | 3fil_B | 3lbl_A |
| 1brt_A | 1jhg_A | 1qu1_D | 1wpu_A | 2cmj_B | 2is8_A | 2v5i_A | 3a8u_X | 3fiq_A | 3lbm_B |
| 1bs3_B | 1jhj_A | 1qve_B | 1wq8_A | 2cmt_A | 2is9_A | 2v5j_A | 3a9b_A | 3fju_B | 3lcc_A |
| 1bs9_A | 1jhs_A | 1qw9_B | 1wqj_B | 2cn3_B | 2it1_A | 2v5z_A | 3a9f_A | 3fkb_E | 3lcm_A |
| 1bsg_A | 1ji1_A | 1qwd_A | 1wqj_I | 2cnz_A | 2iu5_A | 2v6a_O | 3a9j_A | 3fkc_A | 3ld3_A |
| 1bue_A | 1jid_A | 1qwg_A | 1wr8_B | 2cov_I | 2ium_A | 2v6k_B | 3a9l_B | 3fke_A | 3ldd_A |
| 1bx4_A | 1jif_A | 1qwk_A | 1wrd_A | 2cs7_A | 2ivf_A | 2v6u_A | 3a9q_N | 3flg_A | 3le0_A |
| 1bx7_A | 1jke_C | 1qwm_B | 1wri_A | 2cu5_A | 2ivf_B | 2v7w_C | 3a9s_B | 3flv_A | 3le3_A |
| 1bxu_A | 1jkg_A | 1qwz_A | 1wrm_A | 2cvd_D | 2ivn_A | 2v84_A | 3aa0_A | 3fn5_A | 3le4_A |
| 1bxy_A | 1jkx_A | 1qxy_A | 1ws8_A | 2cve_A | 2ivx_A | 2v89_A | 3aa6_B | 3fp5_A | 3let_A |
| 1byi_A | 1jl1_A | 1qy6_A | 1wst_A | 2cvi_B | 2ivy_A | 2v8i_A | 3aaf_B | 3fpc_A | 3lf6_B |
| 1c02_A | 1jl7_A | 1qz9_A | 1wt6_A | 2cwd_A | 2iw0_A | 2v8u_A | 3aal_A | 3fpf_A | 3lfh_F |
| 1c0p_A | 1jlj_A | 1r0r_E | 1wta_A | 2cwi_B | 2iw1_A | 2v9m_A | 3aam_A | 3fpk_B | 3lfj_B |
| 1c1d_A | 1jlt_A | 1r12_A | 1wte_A | 2cwr_A | 2iw2_B | 2v9t_B | 3ab6_A | 3fpr_D | 3lfk_C |
| 1c1k_A | 1jlt_B | 1r17_A | 1wtj_A | 2cws_A | 2iwk_A | 2v9v_A | 3aba_A | 3fpu_B | 3lg5_A |
| 1c1l_A | 1jm1_A | 1r1p_A | 1wto_A | 2cxn_B | 2iwz_A | 2vac_A | 3abf_E | 3fpw_A | 3lgi_A |
| 1c1y_A | 1jnr_C | 1r1t_B | 1wu9_B | 2cyg_A | 2ix4_B | 2vap_A | 3aci_A | 3fq3_C | 3lgn_A |
| 1c1y_B | 1jnr_D | 1r26_A | 1wui_S | 2cz4_A | 2ixc_A | 2vb1_A | 3act_B | 3fqm_A | 3lhq_A |
| 1c4q_B | 1jo0_A | 1r29_A | 1wur_B | 2czd_A | 2ixd_B | 2vba_D | 3acx_A | 3frq_A | 3lhr_B |
| 1c52_A | 1jo8_A | 1r2m_A | 1wve_D | 2czq_B | 2ixk_A | 2vbk_A | 3adg_A | 3frr_A | 3lid_B |
| 1c5e_A | 1jpe_A | 1r2r_B | 1wvf_A | 2d0i_B | 2ixm_A | 2vc3_A | 3ado_A | 3fs7_A | 3lim_D |
| 1c75_A | 1jq5_A | 1r3q_A | 1wwz_B | 2d16_A | 2izz_B | 2vc8_A | 3aey_A | 3ft1_C | 3liy_A |
| 1c7j_A | 1jqe_A | 1r45_B | 1wy1_A | 2d1c_A | 2j1s_A | 2ve8_E | 3afm_A | 3ftd_A | 3ljw_B |
| 1c7k_A | 1jr8_A | 1r55_A | 1wy2_B | 2d1x_A | 2j23_A | 2veb_A | 3afv_A | 3fv3_G | 3lke_A |
| 1c7n_F | 1jsd_B | 1r6d_A | 1wyx_B | 2d1y_C | 2j27_A | 2vfk_A | 3ag3_C | 3fv9_G | 3lkt_B |
| 1cc8_A | 1jt2_A | 1r6j_A | 1wz3_A | 2d29_A | 2j2j_F | 2vfq_A | 3ag3_E | 3fvb_B | 3lkt_Q |
| 1ccw_B | 1ju2_A | 1r6x_A | 1wz8_A | 2d37_A | 2j3x_A | 2vg1_B | 3ag7_A | 3fvh_A | 3llb_A |
| 1cf3_A | 1jub_B | 1r77_B | 1wzd_B | 2d3d_A | 2j5g_A | 2vg3_C | 3agn_A | 3fwa_A | 3llu_A |
| 1cg5_A | 1juv_A | 1r7j_A | 1x0c_A | 2d3n_A | 2j5i_F | 2vgp_D | 3ah2_A | 3fwy_A | 3lny_A |
| 1cg5_B | 1jvw_A | 1r87_A | 1x0l_A | 2d4n_A | 2j5y_A | 2vha_B | 3ahc_A | 3fx4_A | 3log_C |
| 1chd_A | 1jwq_A | 1r88_B | 1x1i_A | 2d4p_A | 2j5z_C | 2vi8_A | 3ahn_A | 3fx7_A | 3lp6_C |
| 1cip_A | 1jy2_N | 1r89_A | 1x1n_A | 2d4v_C | 2j6a_A | 2vig_A | 3ahx_D | 3fy1_B | 3lpc_A |
| 1cjc_A | 1jy2_R | 1r8h_D | 1x1o_B | 2d5b_A | 2j6b_A | 2vj0_A | 3ahy_A | 3fy3_A | 3lpe_B |
| 1cjw_A | 1jy3_P | 1r8s_A | 1x2i_A | 2d5c_A | 2j6f_A | 2vjp_B | 3ahz_A | 3fym_A | 3lpe_G |
| 1cka_A | 1jyh_A | 1r9d_A | 1x2t_C | 2d5k_C | 2j6i_A | 2vjv_B | 3ai3_C | 3fza_A | 3lpw_B |
| 1clc_A | 1jyo_B | 1r9h_A | 1x38_A | 2d5w_B | 2j6l_F | 2vk8_C | 3aia_A | 3fzy_B | 3lqw_A |
| 1cnv_A | 1k07_A | 1r9l_A | 1x3o_A | 2d5z_B | 2j73_B | 2vkj_A | 3aj7_A | 3g00_A | 3lr4_A |
| 1cnz_B | 1k0i_A | 1ra0_A | 1x3x_B | 2d68_B | 2j7j_A | 2vkl_A | 3ajo_A | 3g0e_A | 3lrt_A |
| 1coj_A | 1k0m_A | 1rc9_A | 1x46_A | 2d69_A | 2j7z_A | 2vkv_A | 3ajx_C | 3g0m_A | 3ls0_A |
| 1cpq_A | 1k1e_K | 1rdo_2 | 1x54_A | 2d6m_A | 2j8b_A | 2vla_A | 3ak2_B | 3g1l_A | 3ls9_A |
| 1cqm_B | 1k20_A | 1rfs_A | 1x6i_A | 2d73_A | 2j8g_A | 2vlf_B | 3ak8_H | 3g1p_B | 3ltj_A |
| 1cru_B | 1k2e_A | 1rfy_B | 1x6x_X | 2d7t_H | 2j8h_A | 2vlu_A | 3akb_A | 3g1v_A | 3luc_A |
| 1cs6_A | 1k4i_A | 1rg8_B | 1x8d_C | 2d7t_L | 2j8k_A | 2vmc_A | 3akh_A | 3g20_B | 3lum_D |
| 1ctj_A | 1k4m_C | 1rgx_C | 1x91_A | 2d81_A | 2j8m_A | 2vn4_A | 3alf_A | 3g21_A | 3lvf_P |
| 1cuo_A | 1k5c_A | 1rgz_A | 1x9i_A | 2d8d_B | 2j8w_A | 2vn6_A | 3alu_A | 3g2b_A | 3lw6_A |
| 1cxy_A | 1k5n_A | 1rh6_B | 1x9u_A | 2dc1_B | 2j9c_B | 2vn6_B | 3amn_B | 3g2s_B | 3lwg_B |
| 1cyd_D | 1k66_B | 1rh9_A | 1xcr_A | 2dc3_A | 2j9o_B | 2vng_A | 3ans_B | 3g46_B | 3lwz_A |
| 1cyo_A | 1k6a_A | 1rhc_A | 1xd3_C | 2dc4_B | 2j9w_A | 2vnk_C | 3ap9_A | 3g48_A | 3lx3_A |
| 1cz9_A | 1k6d_B | 1rie_A | 1xdw_A | 2ddb_C | 2ja2_A | 2vnl_A | 3apa_A | 3g5k_D | 3lxr_A |
| 1cza_N | 1k7c_A | 1rjd_A | 1xeo_A | 2ddx_A | 2jab_C | 2vnz_X | 3apr_E | 3g5w_C | 3lxr_F |
| 1czf_A | 1k7i_A | 1rkd_A | 1xes_B | 2de3_A | 2jaf_A | 2vo4_A | 3b34_A | 3g6m_A | 3lxy_A |
| 1czn_A | 1k94_A | 1rki_A | 1xfk_A | 2de6_A | 2jb0_A | 2vo8_A | 3b42_A | 3g7n_B | 3ly0_B |
| 1d02_B | 1k9u_B | 1rkq_A | 1xg0_B | 2de6_F | 2jba_B | 2vo9_B | 3b4u_B | 3g7w_A | 3ly7_A |
| 1d0d_A | 1ka1_A | 1rku_A | 1xg2_A | 2dep_A | 2jbv_A | 2voc_A | 3b4w_A | 3g8h_A | 3lz5_A |
| 1d2n_A | 1kaf_A | 1rl0_A | 1xg2_B | 2dfb_A | 2jc4_A | 2voz_A | 3b51_X | 3g98_B | 3lzo_B |
| 1d4o_A | 1kao_A | 1rlh_A | 1xg4_A | 2dfd_C | 2jc5_A | 2vpg_A | 3b5g_B | 3g9m_B | 3lzw_A |
| 1d4t_A | 1kap_P | 1rlk_A | 1xg7_B | 2dg1_C | 2jcb_A | 2vpj_A | 3b5l_B | 3g9x_A | 3m07_A |
| 1d5l_B | 1kaz_A | 1rm6_A | 1xgs_A | 2dga_A | 2jcq_A | 2vq4_A | 3b5m_B | 3ga3_A | 3m0f_A |
| 1d5t_A | 1kdg_B | 1rm6_B | 1xiw_A | 2dge_B | 2jda_A | 2vqg_D | 3b5n_A | 3ga4_A | 3m0j_A |
| 1d7o_A | 1kdj_A | 1rm6_C | 1xiw_C | 2dgk_A | 2jdd_A | 2vri_A | 3b5n_B | 3gad_F | 3m0m_B |
| 1daa_B | 1kdo_B | 1roc_A | 1xiw_H | 2dho_A | 2jdf_A | 2vrs_C | 3b5n_D | 3gah_A | 3m1h_B |
| 1dbf_A | 1keq_B | 1rp0_B | 1xk4_H | 2dkj_A | 2jdk_D | 2vsv_A | 3b5n_K | 3gbe_A | 3m21_D |
| 1dbw_B | 1kew_A | 1rqj_A | 1xk4_I | 2dko_A | 2je6_B | 2vtc_B | 3b64_A | 3gbs_A | 3m3g_A |
| 1dci_C | 1kfw_A | 1rro_A | 1xky_B | 2dm9_B | 2je8_B | 2vuj_A | 3b6i_A | 3gc6_A | 3m4d_A |
| 1deu_A | 1kg2_A | 1rtq_A | 1xkz_B | 2dp6_A | 2jek_A | 2vun_B | 3b76_A | 3gcz_A | 3m5l_A |
| 1dfu_P | 1kgc_D | 1rtt_A | 1xlq_C | 2dp9_A | 2jep_B | 2vuo_A | 3b7e_A | 3gd6_A | 3m5q_A |
| 1dgf_A | 1khi_A | 1ru0_B | 1xm8_A | 2dpf_D | 2jft_A | 2vv6_D | 3b7s_A | 3gd8_A | 3m66_A |
| 1dhn_A | 1khq_A | 1ru4_A | 1xmk_A | 2dqa_A | 2jg6_A | 2vve_A | 3b84_A | 3gdc_A | 3m6b_A |
| 1dj0_B | 1kid_A | 1rv9_A | 1xmp_B | 2dql_A | 2jh1_A | 2vvp_B | 3b8f_C | 3gdl_B | 3m6z_A |
| 1djr_G | 1kjv_A | 1rwh_A | 1xmt_A | 2dr1_B | 2jhf_B | 2vvt_B | 3b8i_E | 3ge3_A | 3m73_A |
| 1dk8_A | 1klx_A | 1rwj_A | 1xng_B | 2dri_A | 2jhq_A | 2vvw_A | 3b8z_B | 3ge3_E | 3m7o_A |
| 1dl5_B | 1km9_A | 1rwr_A | 1xnk_A | 2drm_B | 2ji7_A | 2vw8_A | 3b9c_C | 3gfu_A | 3m7q_B |
| 1dlf_H | 1kms_A | 1rwy_B | 1xo7_B | 2ds2_D | 2jik_A | 2vwf_A | 3b9d_A | 3gg7_A | 3m8j_A |
| 1dlf_L | 1kmt_A | 1rwz_A | 1xoc_A | 2ds5_A | 2jjc_A | 2vwr_A | 3b9w_A | 3gg9_B | 3m8o_H |
| 1dlj_A | 1kng_A | 1rx0_C | 1xov_A | 2dsj_B | 2jjf_A | 2vws_C | 3ba1_A | 3ggw_C | 3m8o_L |
| 1dlw_A | 1knt_A | 1ry9_C | 1xph_A | 2dsn_B | 2jjn_A | 2vx5_A | 3baa_A | 3ggy_A | 3m8t_B |
| 1dly_A | 1koe_A | 1ryi_B | 1xpp_C | 2dsx_A | 2jjs_C | 2vxn_A | 3bal_B | 3gh6_A | 3m8u_A |
| 1dm1_A | 1kol_B | 1ryo_A | 1xqo_A | 2dt4_A | 2jk9_A | 2vxq_A | 3bbb_D | 3gha_A | 3m91_C |
| 1doi_A | 1kop_A | 1ryp_I | 1xre_A | 2dtj_A | 2jkb_A | 2vxt_I | 3bc1_B | 3gip_A | 3m9q_B |
| 1dok_A | 1kp6_A | 1ryp_J | 1xrk_B | 2dtx_A | 2jkg_A | 2vxt_L | 3bc9_A | 3gir_A | 3mab_A |
| 1dp7_P | 1kpt_A | 1ryp_K | 1xs5_A | 2dur_A | 2jkh_A | 2vxy_A | 3bd1_A | 3gk7_B | 3mao_A |
| 1dpj_A | 1kpu_B | 1ryp_Z | 1xso_B | 2dvm_B | 2jl1_A | 2vyo_A | 3bd2_A | 3gkb_C | 3maz_A |
| 1dpt_A | 1kq1_B | 1rzh_H | 1xsz_A | 2dvx_B | 2jlp_A | 2vyq_A | 3bec_A | 3gkj_A | 3mb4_B |
| 1dqg_A | 1kq3_A | 1rzh_L | 1xt5_A | 2dwu_A | 2jlq_A | 2vyw_A | 3beo_A | 3gkm_A | 3mb5_A |
| 1dqi_A | 1kqp_A | 1rzh_M | 1xtt_C | 2dxe_B | 2lis_A | 2vyx_C | 3ber_A | 3gkr_A | 3mbg_B |
| 1dqp_A | 1kqr_A | 1s1d_B | 1xty_B | 2dy0_A | 2mnr_A | 2vzc_A | 3bex_A | 3gkt_A | 3mbk_B |
| 1dqz_A | 1krh_A | 1s2o_A | 1xu1_D | 2dy1_A | 2nml_A | 2vzm_A | 3bf7_B | 3gkv_B | 3mbx_H |
| 1ds1_A | 1krn_A | 1s3c_A | 1xu1_T | 2dyj_B | 2nmx_A | 2w0p_B | 3bfk_B | 3gl0_A | 3mcb_B |
| 1dsx_C | 1ks8_A | 1s57_B | 1xvg_C | 2dyr_K | 2nn8_A | 2w15_A | 3bfo_B | 3glr_A | 3md1_A |
| 1dsz_A | 1ku1_A | 1s5m_A | 1xvg_E | 2dyu_A | 2nnr_A | 2w1j_B | 3bfq_G | 3glv_B | 3md7_A |
| 1dtd_B | 1kug_A | 1s5u_D | 1xvo_A | 2dze_A | 2nnu_A | 2w1p_A | 3bfv_A | 3gmf_A | 3md9_A |
| 1duv_H | 1kvd_C | 1s69_A | 1xvx_A | 2e0q_A | 2no4_A | 2w1r_A | 3bg8_A | 3gmg_A | 3mdm_A |
| 1dxe_B | 1kve_D | 1s9r_A | 1xw6_A | 2e0t_A | 2npt_D | 2w1v_A | 3bgo_P | 3gmi_A | 3mds_B |
| 1dxj_A | 1kw6_B | 1sa3_A | 1xwt_A | 2e11_B | 2nql_B | 2w20_A | 3bh4_A | 3gms_A | 3mdu_A |
| 1dxy_A | 1kwf_A | 1sat_A | 1xwv_A | 2e1n_B | 2ns1_B | 2w2b_B | 3bh7_A | 3gmv_X | 3meb_B |
| 1dys_B | 1kwg_A | 1sau_A | 1xww_A | 2e1z_A | 2nsf_A | 2w2j_A | 3bh7_B | 3gmx_B | 3mf7_A |
| 1dzk_A | 1kzq_A | 1sbp_A | 1xx1_C | 2e27_L | 2nsz_A | 2w2k_A | 3bj1_C | 3gn9_C | 3mgn_B |
| 1e0w_A | 1l2p_A | 1sd5_A | 1xxq_D | 2e2o_A | 2nt4_A | 2w31_B | 3bje_A | 3gne_A | 3mh9_A |
| 1e25_A | 1l2t_B | 1sdi_A | 1xyz_A | 2e3a_A | 2ntp_A | 2w39_A | 3bkb_A | 3gnr_A | 3mhs_B |
| 1e29_A | 1l3p_A | 1sds_A | 1y0h_A | 2e3z_B | 2nug_B | 2w3g_A | 3bkj_H | 3go2_A | 3mhy_C |
| 1e2w_B | 1l5o_A | 1seg_A | 1y0m_A | 2e42_A | 2nuh_A | 2w3j_A | 3bkr_A | 3go6_A | 3mi4_A |
| 1e3d_B | 1l5w_B | 1sen_A | 1y1p_A | 2e4t_A | 2nuk_A | 2w3p_B | 3bkt_A | 3goc_B | 3mil_A |
| 1e4c_P | 1l6r_A | 1sf9_A | 1y1x_A | 2e5f_A | 2nuw_A | 2w3v_A | 3bl6_A | 3goe_A | 3mjo_B |
| 1e4m_M | 1l6w_B | 1sff_C | 1y20_A | 2e5y_B | 2nw2_B | 2w3x_A | 3bmb_B | 3gon_A | 3mjv_B |
| 1e4v_A | 1l7l_A | 1sfs_A | 1y2t_B | 2e6f_A | 2nx0_A | 2w40_A | 3bmw_A | 3gox_B | 3mkh_B |
| 1e59_A | 1l9l_A | 1sg4_C | 1y37_A | 2e6u_X | 2nx4_C | 2w43_A | 3bmx_B | 3gp3_D | 3mlb_A |
| 1e5k_A | 1l9x_A | 1sgw_A | 1y43_B | 2e7u_A | 2nxb_B | 2w47_A | 3bn6_A | 3gp4_B | 3mm5_B |
| 1e5m_A | 1lb6_A | 1sh7_B | 1y4j_B | 2e7z_A | 2nyb_A | 2w4c_A | 3bnj_A | 3gpg_B | 3mm6_A |
| 1e6i_A | 1lc3_A | 1sh8_B | 1y4w_A | 2e85_A | 2nz7_A | 2w4f_A | 3bo6_B | 3gqh_A | 3mmg_A |
| 1e6y_B | 1lc5_A | 1shu_X | 1y51_A | 2e8e_A | 2nzh_A | 2w4i_F | 3bod_A | 3gqj_A | 3mmh_B |
| 1e7l_B | 1lcl_A | 1skz_A | 1y5i_B | 2e9m_A | 2o07_B | 2w50_B | 3boe_A | 3grh_A | 3mmw_D |
| 1e7s_A | 1lcp_A | 1smo_A | 1y60_E | 2e9y_B | 2o0b_A | 2w5q_A | 3boi_A | 3grn_A | 3mn1_C |
| 1e9g_A | 1le6_A | 1sn2_B | 1y66_C | 2ea3_A | 2o1c_B | 2w70_A | 3bom_C | 3gru_A | 3mos_A |
| 1eaj_B | 1lgt_A | 1snn_A | 1y6i_A | 2eab_A | 2o1k_B | 2w7n_A | 3bom_D | 3gsh_A | 3moy_A |
| 1eao_A | 1lj5_A | 1stm_B | 1y7p_B | 2ebb_A | 2o20_F | 2w7w_B | 3bov_A | 3gt5_A | 3mpc_A |
| 1ear_A | 1lj9_B | 1svb_A | 1y7t_B | 2ebo_C | 2o28_A | 2w83_C | 3bp5_A | 3gv6_A | 3mpz_B |
| 1eb6_A | 1ljo_A | 1svd_M | 1y7w_A | 2ecs_A | 2o2c_C | 2w86_A | 3bpj_B | 3gvf_A | 3mq2_B |
| 1eco_A | 1lk5_B | 1svf_B | 1y80_A | 2ecu_A | 2o2p_D | 2w8x_A | 3bpv_A | 3gvg_B | 3mqd_A |
| 1edg_A | 1lkp_A | 1sw5_B | 1y8a_A | 2eeo_B | 2o36_A | 2w8y_B | 3bpw_A | 3gvo_A | 3mqh_D |
| 1edq_A | 1llf_B | 1swy_A | 1y9l_A | 2eey_A | 2o37_A | 2w91_A | 3bpz_D | 3gw9_C | 3ms5_A |
| 1eej_B | 1llm_C | 1sxq_B | 1y9w_B | 2efr_B | 2o4a_A | 2w98_B | 3bqp_B | 3gwa_A | 3msu_B |
| 1egw_B | 1lo6_A | 1sxr_A | 1y9z_B | 2efv_A | 2o4t_A | 2w9h_A | 3br8_A | 3gwc_D | 3msx_B |
| 1eis_A | 1lq9_A | 1sy7_B | 1yac_B | 2egd_B | 2o4v_B | 2w9r_A | 3bs1_A | 3gwh_B | 3mte_A |
| 1ej0_A | 1lqa_A | 1syy_A | 1yar_D | 2egj_A | 2o5f_B | 2wa2_B | 3bs2_A | 3gwi_A | 3mtr_B |
| 1ej2_A | 1lqv_A | 1szh_A | 1yar_N | 2egv_A | 2o5g_A | 2waa_A | 3bs9_B | 3gwk_E | 3mu7_A |
| 1ejb_A | 1lqx_A | 1szn_A | 1yb0_B | 2eh6_A | 2o5u_A | 2wb6_A | 3bsy_B | 3gwm_A | 3muj_B |
| 1ekj_C | 1lr7_A | 1szo_K | 1yb5_A | 2ehg_A | 2o6f_A | 2wbf_X | 3buv_B | 3gwn_A | 3muz_3 |
| 1ekx_A | 1lrh_A | 1t00_A | 1ybi_B | 2ehq_A | 2o6p_A | 2wbs_A | 3bv6_D | 3gx8_A | 3mwc_A |
| 1elk_A | 1ls6_A | 1t06_A | 1ybk_D | 2eht_A | 2o6s_A | 2wc8_B | 3bvk_F | 3gxb_A | 3mwf_A |
| 1elr_A | 1ls9_A | 1t0b_D | 1ybz_A | 2ehz_A | 2o74_F | 2wci_A | 3bwh_A | 3gxr_B | 3mwj_A |
| 1elu_A | 1lst_A | 1t0f_B | 1yd3_A | 2ei5_B | 2o7i_A | 2wcj_A | 3bwu_D | 3gzg_A | 3mx6_B |
| 1elw_A | 1lt1_H | 1t0f_C | 1ydy_A | 2eiq_B | 2o90_A | 2wco_A | 3bx4_A | 3gzh_A | 3mxn_A |
| 1enf_A | 1lua_C | 1t0p_B | 1yfn_C | 2eix_A | 2o9c_A | 2wcr_A | 3bx4_B | 3gzk_A | 3mxn_B |
| 1eo6_B | 1lvw_B | 1t0t_X | 1yfu_A | 2eiy_B | 2o9s_A | 2wcu_A | 3bxe_B | 3gzx_A | 3mxu_A |
| 1ep0_A | 1lw6_E | 1t1g_A | 1yif_A | 2eja_B | 2oaa_B | 2wdc_A | 3by4_A | 3gzx_B | 3myb_A |
| 1eq9_B | 1lw6_I | 1t1j_B | 1yii_A | 2ejn_B | 2obi_A | 2wds_A | 3byb_B | 3h01_A | 3mzv_B |
| 1es5_A | 1lwb_A | 1t1v_B | 1ykd_A | 2ejw_A | 2obl_A | 2wdu_B | 3byp_A | 3h04_A | 3n08_A |
| 1esw_A | 1lwd_B | 1t2d_A | 1yki_B | 2ejx_A | 2ocg_A | 2we5_C | 3bzz_B | 3h09_B | 3n0i_B |
| 1eu3_A | 1ly2_A | 1t2h_B | 1yn3_B | 2ekp_A | 2ode_D | 2wei_A | 3c05_A | 3h0o_A | 3n10_B |
| 1euh_C | 1m0d_C | 1t2w_C | 1yn8_E | 2eky_C | 2odf_E | 2wf6_A | 3c05_D | 3h0u_C | 3n11_A |
| 1euv_A | 1m0s_B | 1t3q_A | 1yn9_A | 2elc_B | 2odk_C | 2wfc_C | 3c0i_A | 3h12_B | 3n1e_B |
| 1euv_B | 1m0u_A | 1t3q_B | 1ynb_C | 2end_A | 2oe3_A | 2wfh_B | 3c1o_A | 3h1g_A | 3n1f_D |
| 1evh_A | 1m15_A | 1t3y_A | 1ynh_B | 2eo4_A | 2oeb_A | 2wfj_A | 3c2u_A | 3h1s_B | 3n1s_M |
| 1ex2_B | 1m1n_E | 1t4b_B | 1ynp_B | 2eq6_B | 2ofc_A | 2wfo_A | 3c3y_B | 3h34_A | 3n22_A |
| 1ext_A | 1m1n_F | 1t61_C | 1yo3_A | 2erf_A | 2ofk_A | 2wfz_A | 3c4s_A | 3h3n_X | 3n2n_E |
| 1eyh_A | 1m1r_A | 1t61_E | 1yoa_A | 2erw_A | 2og1_A | 2wge_A | 3c5a_A | 3h4n_A | 3n37_A |
| 1eyl_A | 1m2d_B | 1t6c_A | 1yoc_B | 2ery_B | 2oh5_A | 2wgp_B | 3c5e_A | 3h4x_A | 3n3s_A |
| 1eyv_A | 1m2h_A | 1t6g_D | 1yon_A | 2esl_A | 2ohw_B | 2wh7_A | 3c5j_A | 3h55_B | 3n4i_B |
| 1ez3_B | 1m2t_B | 1t6u_L | 1yp1_A | 2et1_A | 2oif_B | 2whg_B | 3c5k_A | 3h5i_A | 3n4j_A |
| 1ezg_B | 1m2x_D | 1t6v_N | 1yph_D | 2etb_A | 2oiz_B | 2wi8_A | 3c68_B | 3h5j_B | 3n5a_A |
| 1ezm_A | 1m3u_C | 1t7q_B | 1yph_E | 2etx_A | 2oj6_C | 2wiy_A | 3c6w_A | 3h5l_B | 3n5b_B |
| 1f08_B | 1m40_A | 1t7r_A | 1yq2_C | 2eu7_X | 2okl_B | 2wj5_A | 3c70_A | 3h62_B | 3n72_B |
| 1f0k_B | 1m4i_A | 1t8h_A | 1yqd_A | 2eut_A | 2okm_A | 2wje_A | 3c7f_A | 3h6p_B | 3n79_A |
| 1f0y_B | 1m4j_A | 1t8k_A | 1yqe_A | 2ev1_A | 2okq_A | 2wjn_C | 3c7t_A | 3h6p_C | 3n98_A |
| 1f1m_C | 1m55_A | 1t8t_B | 1yqw_B | 2evb_A | 2ol1_B | 2wjn_L | 3c7x_A | 3h78_A | 3n9g_H |
| 1f1u_A | 1m5t_A | 1t8z_C | 1yqw_Q | 2ewh_A | 2olm_A | 2wjn_M | 3c8e_A | 3h7h_A | 3n9i_B |
| 1f39_A | 1m6j_B | 1t92_B | 1yqz_A | 2ewt_A | 2oln_A | 2wk0_A | 3c8i_A | 3h7h_B | 3n9u_B |
| 1f3u_G | 1m70_D | 1t9i_B | 1yrk_A | 2ex0_B | 2olp_A | 2wkk_C | 3c8o_A | 3h7i_A | 3n9u_C |
| 1f46_B | 1m7a_B | 1ta3_A | 1ys1_X | 2ex2_A | 2olr_A | 2wkx_A | 3c8p_A | 3h7r_A | 3nbk_A |
| 1f4p_A | 1m7g_A | 1ta3_B | 1ysl_B | 2exh_D | 2omy_B | 2wl1_A | 3c97_A | 3h7u_A | 3ncl_A |
| 1f5j_A | 1m7j_A | 1ta9_B | 1yt3_A | 2exv_A | 2omz_A | 2wm3_A | 3c9a_B | 3h81_C | 3ndd_A |
| 1f5v_A | 1m7s_D | 1tag_A | 1ytq_A | 2ez9_A | 2on5_A | 2wm8_A | 3c9h_B | 3h87_B | 3ndh_B |
| 1f60_A | 1m8s_A | 1taw_B | 1yu0_A | 2f0c_A | 2oo1_B | 2wmf_A | 3c9u_B | 3h87_D | 3ndj_A |
| 1f60_B | 1m93_B | 1tbf_A | 1yuz_B | 2f23_B | 2op3_A | 2wn3_C | 3c9x_A | 3h8g_C | 3ndo_A |
| 1f7l_A | 1m9z_A | 1tc5_B | 1yw5_A | 2f2b_A | 2op6_A | 2wnp_F | 3c9z_A | 3h8t_A | 3nfu_A |
| 1f8m_C | 1mb4_A | 1tca_A | 1ywm_A | 2f51_A | 2opc_A | 2wns_A | 3ca7_A | 3h8x_A | 3nfw_B |
| 1f94_A | 1mc2_A | 1ten_A | 1yxy_A | 2f5g_B | 2opg_B | 2wnv_F | 3cai_A | 3h91_A | 3ng7_X |
| 1f9f_D | 1md6_A | 1tez_B | 1yya_A | 2f5t_X | 2oqb_A | 2wnx_A | 3cb0_D | 3h9c_A | 3ngf_A |
| 1fcz_A | 1me4_A | 1tf1_A | 1yzf_A | 2f5v_A | 2or7_A | 2wny_B | 3cb6_A | 3h9e_O | 3ngh_A |
| 1fd3_A | 1mex_L | 1tf4_A | 1yzl_A | 2f60_K | 2os5_A | 2woe_C | 3cbq_A | 3ha9_A | 3ngj_A |
| 1fe0_B | 1mfa_H | 1tg0_A | 1yzm_A | 2f6e_A | 2os9_B | 2wol_A | 3cbx_A | 3hc4_L | 3ngp_A |
| 1fec_B | 1mg4_A | 1tg7_A | 1yzq_A | 2f6u_A | 2osa_A | 2wot_A | 3ccd_B | 3hcj_A | 3ni0_A |
| 1feh_A | 1mgq_C | 1tgr_A | 1yzx_A | 2f7b_A | 2otu_E | 2wp7_A | 3ce6_A | 3hcn_B | 3ni2_A |
| 1fj2_B | 1mgr_A | 1tgx_B | 1z06_A | 2f91_A | 2oui_A | 2wpq_C | 3ce7_A | 3hd4_A | 3nis_B |
| 1fk5_A | 1mgt_A | 1th7_H | 1z08_C | 2f99_C | 2ous_A | 2wpv_D | 3cfc_H | 3hdf_B | 3nj2_A |
| 1flj_A | 1mhn_A | 1thg_A | 1z0j_A | 2f9i_D | 2ov0_A | 2wpv_E | 3cfz_A | 3hdl_A | 3nje_B |
| 1fm0_D | 1mhx_A | 1thm_A | 1z0j_B | 2f9n_B | 2ow9_B | 2wqi_D | 3cg1_A | 3he5_C | 3njn_C |
| 1fm4_A | 1mi3_B | 1thx_A | 1z0s_C | 2fao_B | 2ows_A | 2wqk_A | 3cgi_C | 3he5_D | 3nn1_A |
| 1fn9_A | 1mix_A | 1thz_A | 1z1s_A | 2fb5_A | 2ox0_A | 2wqr_A | 3chj_A | 3he8_B | 3no0_A |
| 1fob_A | 1mj5_A | 1tjy_A | 1z2a_A | 2fba_A | 2ox4_H | 2wsb_C | 3chm_A | 3hef_B | 3no7_A |
| 1fp2_A | 1mk0_A | 1tke_A | 1z2n_X | 2fbd_A | 2ox6_B | 2wt1_A | 3ci7_A | 3hf5_C | 3noj_A |
| 1fpo_B | 1mkk_A | 1tn4_A | 1z2u_A | 2fbn_A | 2oxc_A | 2wta_A | 3cij_A | 3hfo_A | 3nok_A |
| 1fqt_B | 1mla_A | 1to4_A | 1z3e_A | 2fbq_A | 2oxg_A | 2wtg_A | 3cin_A | 3hfw_A | 3nol_A |
| 1fr3_A | 1mlw_A | 1toa_B | 1z3e_B | 2fc3_A | 2oxg_Y | 2wtm_C | 3cip_G | 3hg3_B | 3noo_B |
| 1frb_A | 1mn8_B | 1tov_A | 1z3q_A | 2fcb_A | 2oxn_A | 2wto_A | 3civ_A | 3hgb_A | 3nqx_A |
| 1fsg_A | 1mo0_B | 1tp5_A | 1z47_A | 2fcr_A | 2oy2_F | 2wu9_A | 3cjp_B | 3hgm_A | 3nr1_A |
| 1ftr_A | 1mo9_A | 1tp6_A | 1z4j_A | 2fcw_A | 2oy7_A | 2wue_A | 3cjs_B | 3hgu_A | 3ns2_A |
| 1fus_A | 1mof_A | 1tp9_B | 1z4r_A | 2fcw_B | 2oya_B | 2wuh_A | 3ckf_A | 3hh7_A | 3ns6_B |
| 1fw4_A | 1moq_A | 1tqh_A | 1z6n_A | 2fd5_A | 2oyn_A | 2wuk_A | 3cl5_A | 3hh8_A | 3nsl_F |
| 1fx4_A | 1mpg_B | 1tr0_J | 1z6o_D | 2fdn_A | 2oyp_A | 2wut_A | 3cla_A | 3hhi_B | 3nsx_B |
| 1fxl_A | 1mpl_A | 1tsf_A | 1z6o_M | 2fdv_A | 2ozf_A | 2wv3_A | 3cls_D | 3hht_B | 3nsz_A |
| 1fxo_G | 1mug_A | 1tt2_A | 1z76_B | 2fdx_A | 2ozl_A | 2wvf_A | 3cm0_A | 3his_A | 3nt1_B |
| 1fz1_B | 1mv8_C | 1tt8_A | 1z7x_W | 2fe3_A | 2ozn_A | 2wvg_F | 3cmj_A | 3hjb_A | 3ntk_A |
| 1g01_A | 1mvf_A | 1tu1_B | 1z8o_A | 2fe5_A | 2ozn_B | 2wvv_A | 3cmy_A | 3hje_A | 3nua_A |
| 1g0o_C | 1mvf_D | 1tu7_B | 1z96_A | 2fe8_A | 2p02_A | 2ww2_B | 3cnk_B | 3hjr_A | 3nv1_A |
| 1g12_A | 1mvo_A | 1tu9_A | 1zbf_A | 2ff4_A | 2p09_A | 2ww5_A | 3cnm_A | 3hjs_A | 3nvs_A |
| 1g1s_B | 1mxg_A | 1tua_A | 1zch_A | 2ffu_A | 2p0f_A | 2wwe_A | 3cnu_A | 3hl5_B | 3nvw_J |
| 1g1t_A | 1mxr_A | 1tuh_A | 1zcz_A | 2ffy_B | 2p1g_B | 2wwf_C | 3cou_A | 3hlx_A | 3nwo_A |
| 1g29_2 | 1my5_A | 1tuk_A | 1zd0_A | 2fgo_A | 2p1m_B | 2wwk_T | 3cp5_A | 3hm5_A | 3nxb_B |
| 1g2o_C | 1my6_B | 1tvn_B | 1zd8_A | 2fgr_A | 2p39_A | 2wwx_A | 3cp7_B | 3hmc_A | 3ny3_A |
| 1g2q_A | 1mz4_A | 1txg_B | 1zdy_A | 2fhf_A | 2p3h_A | 2wwx_B | 3cpq_B | 3hms_A | 3ny7_A |
| 1g2r_A | 1mz9_A | 1ty9_A | 1zem_A | 2fhz_A | 2p49_B | 2wx9_A | 3cpt_A | 3hnb_M | 3nye_A |
| 1g3k_B | 1mzy_A | 1tzw_A | 1zgd_B | 2fhz_B | 2p4e_P | 2wy3_B | 3cq5_B | 3hnx_A | 3nyk_A |
| 1g4i_A | 1n08_B | 1tzy_C | 1zhh_A | 2fi1_A | 2p4k_A | 2wy4_A | 3cql_A | 3hol_A | 3nyt_A |
| 1g5a_A | 1n0q_B | 1tzy_F | 1zhq_A | 2fi9_A | 2p51_A | 2wy7_Q | 3cqt_A | 3hpc_X | 3nzn_B |
| 1g61_A | 1n13_G | 1u07_A | 1zhv_A | 2fj8_A | 2p54_A | 2wy8_A | 3cry_A | 3hr0_A | 3o07_A |
| 1g6a_A | 1n1j_A | 1u09_A | 1zhx_A | 2fl4_A | 2p57_A | 2wya_A | 3csk_A | 3hra_A | 3o0a_B |
| 1g6c_B | 1n1j_B | 1u11_A | 1zi9_A | 2flh_B | 2p5k_A | 2wyq_A | 3ct1_A | 3hrx_A | 3o0d_C |
| 1g6g_A | 1n3y_A | 1u1w_B | 1zja_A | 2fli_A | 2p5y_A | 2wz9_A | 3ctg_A | 3hs3_A | 3o0g_D |
| 1g6h_A | 1n5w_D | 1u2b_A | 1zjc_A | 2fm6_A | 2p65_A | 2wzm_A | 3ctk_A | 3hsh_A | 3o1g_A |
| 1g6u_B | 1n63_C | 1u2h_A | 1zjj_B | 2fma_A | 2p68_A | 2wzx_A | 3ctp_B | 3hss_B | 3o1k_B |
| 1g8a_A | 1n63_E | 1u2p_A | 1zk4_A | 2fmm_C | 2p6h_B | 2x0k_A | 3ctz_A | 3ht1_A | 3o1n_A |
| 1g8i_B | 1n71_C | 1u2w_B | 1zk7_A | 2fmp_A | 2p6w_A | 2x18_E | 3cu4_A | 3ht5_A | 3o1p_A |
| 1g8k_A | 1n7e_A | 1u3i_A | 1zkc_A | 2fn3_A | 2p6z_A | 2x1b_A | 3cu9_A | 3huh_A | 3o26_A |
| 1g8k_D | 1n7s_A | 1u53_A | 1zke_B | 2fn4_A | 2p74_A | 2x23_B | 3cui_A | 3hup_B | 3o2e_A |
| 1g94_A | 1n7s_B | 1u55_A | 1zkk_B | 2fn9_B | 2p8b_A | 2x2o_A | 3cvb_A | 3hv2_B | 3o3m_C |
| 1g97_A | 1n7s_C | 1u5d_B | 1zkl_A | 2fne_A | 2p9x_D | 2x32_A | 3cwn_B | 3hvi_A | 3o3m_D |
| 1g9g_A | 1n7s_D | 1u5f_A | 1zl0_B | 2fnu_A | 2pa1_A | 2x3h_C | 3cwv_A | 3hvu_C | 3o3u_N |
| 1g9o_A | 1n82_B | 1u60_B | 1zlm_A | 2fo3_A | 2pa6_B | 2x3m_A | 3cx5_E | 3hvv_A | 3o4h_A |
| 1ga6_A | 1n83_A | 1u6e_A | 1zm8_A | 2fp1_A | 2pa7_B | 2x49_A | 3cx5_F | 3hx9_A | 3o4r_B |
| 1gai_A | 1n8f_B | 1u6r_B | 1zn8_A | 2fp7_A | 2pbc_C | 2x4d_A | 3cx5_G | 3hxa_F | 3o4v_B |
| 1gbg_A | 1n9l_A | 1u6t_A | 1zo2_A | 2fq3_A | 2pbd_P | 2x4k_B | 3cx5_I | 3hxs_B | 3o5v_B |
| 1gbs_A | 1na0_A | 1u6z_B | 1zoi_B | 2fqw_A | 2pbi_C | 2x4l_A | 3cx5_O | 3hxw_A | 3o70_A |
| 1gci_A | 1na5_A | 1u84_A | 1zos_C | 2fr2_A | 2pbi_D | 2x5c_B | 3cxk_A | 3hz2_A | 3o79_A |
| 1gcq_C | 1nb9_A | 1u8f_Q | 1zps_B | 2fr5_C | 2pbp_A | 2x5f_B | 3cxz_A | 3hzb_E | 3o7b_A |
| 1gde_B | 1nbc_B | 1u8v_C | 1zpw_X | 2frg_P | 2pc8_A | 2x5h_B | 3cy4_A | 3i0w_A | 3o83_B |
| 1gee_E | 1nbu_D | 1u8y_B | 1zq9_B | 2ft0_A | 2pcj_B | 2x5x_A | 3cyi_A | 3i1a_A | 3o85_B |
| 1geg_G | 1nc5_A | 1u9k_A | 1zr0_D | 2ftx_B | 2pcn_A | 2x5y_A | 3cz1_B | 3i1u_A | 3o8m_A |
| 1ges_B | 1ndd_A | 1ua4_A | 1zr3_B | 2fu0_A | 2pdr_B | 2x6w_A | 3czf_B | 3i24_A | 3o9z_A |
| 1gk6_A | 1ne7_C | 1ua6_L | 1zr6_A | 2fu4_A | 2pfz_A | 2x7b_A | 3czt_X | 3i26_D | 3oa3_B |
| 1gk7_A | 1nep_A | 1uai_A | 1zs4_D | 2fuk_A | 2pg0_B | 2x7k_A | 3czz_B | 3i2z_A | 3oaj_A |
| 1gk9_A | 1nf8_A | 1uas_A | 1zsw_A | 2fvh_A | 2pgo_A | 2x7m_A | 3d03_B | 3i31_A | 3oam_A |
| 1gk9_B | 1nff_B | 1ub3_A | 1zsx_A | 2fvv_A | 2ph3_A | 2x8h_A | 3d0n_A | 3i33_A | 3obu_A |
| 1gl2_A | 1nfv_N | 1uc4_A | 1zt5_A | 2fvy_A | 2phn_A | 2x8r_A | 3d0o_A | 3i35_A | 3ocu_A |
| 1gl2_B | 1ng6_A | 1uc4_G | 1zu3_A | 2fwh_A | 2pi6_A | 2x8s_A | 3d1b_C | 3i36_A | 3od9_A |
| 1gl2_C | 1nh2_A | 1uca_A | 1zuo_A | 2fyg_A | 2pie_A | 2x8x_X | 3d1g_A | 3i3f_B | 3odg_A |
| 1gl2_D | 1nh2_B | 1ucr_B | 1zuu_A | 2fyx_A | 2piy_B | 2x96_A | 3d1k_A | 3i3g_A | 3ofk_C |
| 1gmu_C | 1nhc_E | 1ucs_A | 1zuy_A | 2fzp_A | 2pjz_A | 2xb4_A | 3d2q_A | 3i45_A | 3og9_B |
| 1gmy_A | 1nhk_L | 1udc_A | 1zv1_A | 2fzv_B | 2pk3_A | 2xbk_A | 3d2w_A | 3i47_A | 3ogn_B |
| 1gn0_A | 1nki_B | 1ueb_A | 1zwh_A | 2fzw_B | 2pk8_A | 2xbl_A | 3d30_A | 3i48_B | 3ogr_A |
| 1gnl_A | 1nkp_D | 1uek_A | 1zwz_A | 2g2n_C | 2pkf_A | 2xbp_A | 3d32_A | 3i4o_B | 3oid_C |
| 1gnt_A | 1nln_A | 1uf5_A | 1zx6_A | 2g2s_A | 2pko_A | 2xc2_A | 3d34_A | 3i4s_A | 3oig_A |
| 1gny_A | 1nls_A | 1ufb_C | 1zxt_B | 2g30_A | 2pkt_A | 2xcb_A | 3d3b_A | 3i4z_B | 3oiu_A |
| 1go3_N | 1nnf_A | 1ufi_B | 1zxx_A | 2g45_D | 2plt_A | 2xce_F | 3d3z_A | 3i57_B | 3oj7_A |
| 1goi_B | 1nnh_A | 1ufy_A | 1zz0_A | 2g5x_A | 2pmk_A | 2xcj_A | 3d43_B | 3i5c_B | 3ojs_A |
| 1gp6_A | 1nns_A | 1ug6_A | 1zzg_B | 2g64_A | 2pmr_A | 2xcz_A | 3d47_A | 3i5r_A | 3ol0_A |
| 1gpe_A | 1nnw_B | 1ugi_E | 1zzk_A | 2g6f_X | 2pn6_A | 2xda_A | 3d4i_A | 3i5x_A | 3ol3_A |
| 1gpi_A | 1noa_A | 1ugp_B | 1zzo_A | 2g76_A | 2pn8_B | 2xde_A | 3d4u_B | 3i6c_A | 3omc_B |
| 1gpu_A | 1nof_A | 1ugx_A | 1zzw_A | 2g7o_A | 2pnd_A | 2xdg_A | 3d6r_A | 3i6t_B | 3omt_A |
| 1gq1_A | 1nog_A | 1uha_A | 256b_B | 2g84_A | 2pnx_A | 2xdj_F | 3d79_A | 3i7u_B | 3onr_I |
| 1gq8_A | 1nox_A | 1uhe_A | 2a07_K | 2g8o_B | 2pny_A | 2xdp_A | 3d7a_B | 3i8s_C | 3oo8_A |
| 1gql_B | 1npy_B | 1uhk_B | 2a0n_A | 2g9f_A | 2poi_A | 2xdw_A | 3d8t_A | 3i94_A | 3ooi_A |
| 1gqv_A | 1nq6_A | 1ui0_A | 2a14_A | 2ga4_D | 2pok_B | 2xe4_A | 3d95_B | 3i96_A | 3op8_A |
| 1gtf_I | 1nq7_A | 1uiw_C | 2a15_A | 2gag_A | 2por_A | 2xed_A | 3d9t_A | 3i98_E | 3opk_A |
| 1gtv_A | 1nqu_B | 1uj0_A | 2a26_C | 2gag_B | 2pos_D | 2xet_B | 3d9x_C | 3i9q_A | 3oq2_A |
| 1gtz_D | 1nr0_A | 1uj2_A | 2a28_A | 2gag_C | 2ppp_A | 2xeu_A | 3d9y_A | 3ia2_F | 3oqy_B |
| 1gu2_A | 1nr4_G | 1uj8_A | 2a2n_C | 2gas_A | 2ppt_B | 2xev_A | 3da0_C | 3ia4_D | 3orh_C |
| 1gu7_B | 1nsc_B | 1uk7_A | 2a2r_B | 2gb4_B | 2pqm_B | 2xf2_A | 3dac_A | 3ian_A | 3orv_B |
| 1gud_A | 1nth_A | 1ukf_A | 2a40_E | 2gbt_A | 2pqr_B | 2xfd_A | 3dai_A | 3iar_A | 3orv_D |
| 1gug_D | 1ntv_A | 1ukm_A | 2a4v_A | 2gbw_E | 2pqr_D | 2xfv_A | 3dan_A | 3iav_A | 3ose_A |
| 1gui_A | 1nty_A | 1ukm_B | 2a4x_A | 2gc4_L | 2pqx_A | 2xh2_C | 3daq_A | 3ib7_A | 3osm_A |
| 1gv5_A | 1nvm_C | 1uku_A | 2a53_C | 2gdq_B | 2pr5_A | 2xhi_A | 3dau_A | 3ich_A | 3oti_B |
| 1gvj_A | 1nvm_F | 1ukz_A | 2a5d_A | 2gdz_A | 2psd_A | 2xhn_A | 3dc5_C | 3id1_A | 3ouf_B |
| 1gvn_D | 1nw2_H | 1ulk_A | 2a61_C | 2gec_A | 2psp_B | 2xi8_A | 3dcn_A | 3id7_A | 3ovp_B |
| 1gwe_A | 1nwa_A | 1ulr_A | 2a6s_B | 2gey_D | 2pst_X | 2xij_A | 3del_B | 3ida_A | 3oyy_B |
| 1gwi_B | 1nwp_A | 1umd_C | 2a6x_A | 2gf3_B | 2pth_A | 2xio_A | 3deo_A | 3idw_A | 3p0t_A |
| 1gwu_A | 1nww_A | 1umd_D | 2a70_A | 2gf9_A | 2ptt_B | 2xkg_A | 3dfg_A | 3ie4_A | 3p1f_A |
| 1gxn_A | 1nwz_A | 1umk_A | 2a7l_B | 2gg6_A | 2ptz_A | 2xkr_A | 3dgb_A | 3ie5_A | 3p1g_A |
| 1gxu_A | 1nxc_A | 1umz_A | 2a8n_A | 2gh0_B | 2pu3_A | 2xla_A | 3dgp_A | 3iei_C | 3p2n_A |
| 1gxy_A | 1nxj_A | 1uow_A | 2a9i_A | 2gh0_D | 2pu9_A | 2xlk_A | 3dgp_B | 3iev_A | 3p2t_A |
| 1gy6_A | 1nxm_A | 1uoy_A | 2a9s_A | 2gh9_A | 2pu9_B | 2xm5_A | 3dgt_A | 3iez_B | 3p3c_A |
| 1gy7_C | 1nyk_B | 1uqx_A | 2aa1_B | 2gha_A | 2pv2_A | 2xmx_A | 3dha_A | 3ife_A | 3p3e_A |
| 1gyh_C | 1nyt_C | 1uqz_A | 2aal_C | 2gia_B | 2pvb_A | 2xn4_A | 3dhi_B | 3ig9_C | 3p3g_A |
| 1gyo_A | 1nza_A | 1urn_C | 2aan_A | 2gib_B | 2pve_B | 2xn6_A | 3dhi_C | 3igz_B | 3p3o_A |
| 1gyv_A | 1o04_E | 1urr_A | 2ab0_A | 2giy_A | 2pvq_A | 2xov_A | 3dho_C | 3ihw_A | 3p48_A |
| 1gyy_B | 1o0e_B | 1urs_A | 2abk_A | 2gj3_A | 2pwy_A | 2xpp_A | 3die_A | 3ihz_B | 3p4t_A |
| 1gzc_A | 1o1z_A | 1us5_A | 2abw_B | 2gjd_C | 2pxx_A | 2xqu_A | 3dj9_A | 3ii7_A | 3p5h_A |
| 1gzw_B | 1o26_C | 1use_A | 2acf_D | 2gke_A | 2py4_A | 2xs4_A | 3djh_C | 3iij_A | 3p73_A |
| 1h03_P | 1o4k_A | 1usf_B | 2ad6_A | 2gkm_B | 2pyw_A | 2xsu_A | 3djl_A | 3iiu_M | 3p73_B |
| 1h0h_B | 1o4s_A | 1usg_A | 2ad6_D | 2gkr_I | 2pz0_B | 2xt2_A | 3djo_A | 3ij3_A | 3p7y_A |
| 1h16_A | 1o4t_A | 1uso_A | 2ae2_A | 2gl5_A | 2pze_B | 2xts_A | 3dk9_A | 3ijl_A | 3p97_C |
| 1h1n_A | 1o4v_A | 1usq_B | 2aen_E | 2gmy_E | 2pzh_B | 2xts_B | 3dkc_A | 3ik7_D | 3p9c_A |
| 1h1y_A | 1o4y_A | 1uti_A | 2aex_A | 2gn4_B | 2q0l_A | 2xtt_B | 3dkm_A | 3ilo_A | 3p9p_A |
| 1h2b_B | 1o5u_A | 1uu4_A | 2ag4_B | 2gnc_A | 2q20_B | 2xu3_A | 3dkr_A | 3ils_A | 3p9x_A |
| 1h2c_A | 1o5x_A | 1uuq_A | 2ag5_B | 2gok_A | 2q28_A | 2xu8_B | 3dl0_A | 3ilw_A | 3pb6_X |
| 1h2e_A | 1o7e_B | 1uuy_A | 2agd_B | 2gom_A | 2q2a_D | 2xvm_B | 3dlm_A | 3im1_A | 3pbf_A |
| 1h2s_A | 1o7i_A | 1uv4_A | 2ahf_A | 2gou_A | 2q2h_A | 2xvs_A | 3dm8_A | 3im9_A | 3pc3_A |
| 1h2s_B | 1o7j_C | 1uvq_A | 2ahn_A | 2gpe_B | 2q35_A | 2xvx_A | 3dme_B | 3imh_A | 3pcv_A |
| 1h4a_X | 1o7q_B | 1uw4_C | 2aib_A | 2gqt_A | 2q5c_A | 2xws_A | 3dmg_A | 3inz_B | 3pd2_B |
| 1h4g_A | 1o7z_B | 1uw4_D | 2akz_B | 2gqw_A | 2q62_G | 2xwt_C | 3dmi_A | 3iof_A | 3pd7_A |
| 1h4p_A | 1o82_A | 1uwc_A | 2anv_A | 2grc_A | 2q73_C | 2xxj_D | 3dmo_A | 3ioh_A | 3pdn_A |
| 1h4r_A | 1o8s_A | 1uwf_A | 2any_A | 2grr_B | 2q86_B | 2xxl_B | 3dnf_B | 3ioq_A | 3pel_B |
| 1h5b_B | 1o8x_A | 1uwk_B | 2ap1_A | 2gsd_A | 2q87_A | 2xy2_A | 3dpg_B | 3iox_A | 3pew_A |
| 1h5q_L | 1o91_C | 1uwz_A | 2apg_A | 2gso_B | 2q88_A | 2xz2_A | 3dqg_A | 3ip4_A | 3pf2_A |
| 1h5v_A | 1o98_A | 1uxx_X | 2aqm_A | 2gte_A | 2q8n_C | 2xzi_A | 3dr0_C | 3ip8_A | 3pfg_A |
| 1h64_Q | 1o9i_D | 1uxy_A | 2aqp_A | 2gtr_A | 2q8r_G | 2y2z_A | 3dr4_B | 3ipc_A | 3pfs_A |
| 1h6f_A | 1o9r_E | 1uy1_A | 2ar1_A | 2gu3_A | 2q9u_A | 2y39_A | 3dra_A | 3ipf_A | 3pg6_C |
| 1h6l_A | 1oa2_C | 1uyx_A | 2arc_B | 2gud_B | 2qa9_E | 2y3q_B | 3drf_A | 3ipw_A | 3pgx_A |
| 1h6u_A | 1oa8_A | 1uz3_A | 2asd_A | 2gui_A | 2qac_A | 2y3v_D | 3drw_B | 3iq3_A | 3phs_A |
| 1h6w_A | 1oaa_A | 1v05_A | 2asu_B | 2guv_C | 2qap_A | 2y3z_A | 3drz_B | 3iql_A | 3phx_B |
| 1h72_C | 1oai_A | 1v08_B | 2at8_X | 2guy_A | 2qb7_A | 2y5p_C | 3ds4_B | 3irp_X | 3pjp_B |
| 1h75_A | 1oal_A | 1v0z_B | 2atb_A | 2gw4_D | 2qc5_A | 2y7b_A | 3dsk_A | 3irs_A | 3pk0_A |
| 1h7e_B | 1oao_A | 1v2z_A | 2atv_A | 2gwm_A | 2qd6_A | 2y88_A | 3dso_A | 3irv_A | 3pkv_A |
| 1h8p_B | 1oaq_H | 1v30_A | 2au7_A | 2gxg_A | 2qdx_A | 2y8m_A | 3dt9_A | 3is3_A | 3plf_D |
| 1h8u_A | 1oaq_L | 1v33_A | 2avd_B | 2gyq_B | 2qed_A | 2yay_A | 3dtb_A | 3isa_B | 3plw_A |
| 1h97_B | 1obo_A | 1v37_A | 2avk_A | 2gz1_B | 2qee_F | 2ygs_A | 3dvw_A | 3iso_A | 3plx_B |
| 1h98_A | 1oc2_B | 1v4p_C | 2axq_A | 2gz4_A | 2qev_A | 2yqu_B | 3dwg_A | 3isq_A | 3pmc_B |
| 1h9m_A | 1oc8_A | 1v4x_B | 2axw_B | 2gze_A | 2qf4_B | 2yrr_B | 3dwg_C | 3it4_B | 3pmd_A |
| 1h9s_B | 1ock_A | 1v54_A | 2ayd_A | 2gze_B | 2qfa_A | 2ysk_A | 3dwv_B | 3it4_C | 3pms_A |
| 1hbn_C | 1ocy_A | 1v54_J | 2b0a_A | 2gzg_B | 2qfa_B | 2yva_B | 3dxt_A | 3iu5_A | 3pmt_A |
| 1hbn_E | 1odm_A | 1v54_V | 2b0t_A | 2h17_A | 2qfa_C | 2yve_A | 3dy0_A | 3iu7_A | 3pna_A |
| 1hc9_B | 1odt_H | 1v55_D | 2b1k_A | 2h1c_A | 2qfe_A | 2yvi_A | 3dzw_A | 3iux_A | 3po0_A |
| 1hd2_A | 1oe2_A | 1v55_L | 2b3f_D | 2h1v_A | 2qg1_A | 2yvo_A | 3e05_B | 3iwt_A | 3po8_A |
| 1hdo_A | 1off_A | 1v58_B | 2b3h_A | 2h2b_A | 2qgy_B | 2yvt_A | 3e0i_A | 3ix3_B | 3pqa_B |
| 1hfe_L | 1ofl_A | 1v5d_A | 2b49_A | 2h2r_B | 2qhl_B | 2yw2_A | 3e13_X | 3ixq_D | 3pr9_A |
| 1hfe_T | 1ofs_C | 1v5f_A | 2b4z_A | 2h2z_A | 2qho_B | 2yw3_A | 3e17_B | 3jpz_B | 3prp_A |
| 1hfo_E | 1ofs_D | 1v5i_B | 2b5a_A | 2h3h_A | 2qhs_A | 2ywd_A | 3e2d_A | 3jqj_C | 3psm_A |
| 1hfs_A | 1ofw_A | 1v6s_A | 2b5h_A | 2h3l_A | 2qia_A | 2ywj_A | 3e3u_A | 3jql_A | 3pua_A |
| 1hh8_A | 1ofz_A | 1v70_A | 2b5w_A | 2h54_B | 2qif_A | 2ywk_A | 3e4g_A | 3jqu_A | 3pvi_B |
| 1hj8_A | 1ogd_D | 1v7p_C | 2b6n_A | 2h62_A | 2qih_B | 2yxm_A | 3e4w_B | 3jqy_C | 3pxl_A |
| 1hjs_B | 1ogm_X | 1v7r_A | 2b7r_A | 2h64_B | 2qim_A | 2yxn_A | 3e55_A | 3jr0_B | 3q0h_A |
| 1hl7_B | 1oh0_A | 1v7z_F | 2b82_A | 2h6f_A | 2qkh_A | 2yxo_B | 3e6j_A | 3js4_B | 3q12_C |
| 1hle_A | 1oh4_A | 1v8c_C | 2b9w_A | 2h6f_B | 2qkp_C | 2yxw_A | 3e6s_F | 3js5_A | 3q1x_A |
| 1hlq_C | 1oh9_A | 1v8f_A | 2ba2_C | 2h6n_A | 2qlt_A | 2yxz_A | 3e6z_X | 3js8_A | 3q20_B |
| 1hm6_B | 1ohp_B | 1v8h_A | 2bay_E | 2h6u_G | 2qmc_A | 2yyv_B | 3e7d_A | 3jsl_B | 3q23_B |
| 1hml_A | 1oi6_B | 1v93_A | 2bba_A | 2h88_A | 2qmc_D | 2yyy_A | 3e7h_A | 3jsy_B | 3q2e_A |
| 1hmt_A | 1ojk_A | 1v96_B | 2bbe_A | 2h88_B | 2qmm_A | 2yz1_B | 3e7r_L | 3jte_A | 3q3u_A |
| 1hnj_A | 1ojq_A | 1v98_A | 2bcg_G | 2h88_D | 2qmq_A | 2yzc_D | 3e8m_B | 3jtm_A | 3q49_B |
| 1hp1_A | 1ojx_C | 1v9f_A | 2bcg_Y | 2h88_P | 2qn0_A | 2yzh_C | 3e8t_A | 3jtz_A | 3q4t_A |
| 1hpg_A | 1ok0_A | 1vbi_A | 2bcm_B | 2h8e_A | 2qnw_A | 2yzt_A | 3e96_B | 3ju2_A | 3q4u_A |
| 1hq0_A | 1okt_A | 1vbu_A | 2bcr_A | 2h8g_B | 2qo4_A | 2z0a_B | 3e9t_B | 3juu_A | 3q5y_A |
| 1ht6_A | 1on3_A | 1vbw_A | 2bd0_D | 2h9b_A | 2qpn_B | 2z0j_E | 3ea3_B | 3jva_F | 3q62_B |
| 1ht9_B | 1ong_A | 1vc4_B | 2bek_D | 2h9h_A | 2qpw_A | 2z0m_A | 3ea6_A | 3jxo_A | 3q6d_B |
| 1hw1_B | 1oni_C | 1vcd_A | 2bem_A | 2ha8_B | 2qq4_B | 2z0t_C | 3eaz_A | 3jxs_A | 3q6l_A |
| 1hx0_A | 1onj_A | 1vcl_A | 2beq_D | 2hax_B | 2qqi_A | 2z0x_A | 3ebh_A | 3jxy_A | 3q8g_A |
| 1hx1_B | 1ooe_A | 1vd1_A | 2bez_C | 2haz_A | 2qrl_A | 2z1a_A | 3ec0_B | 3jyo_A | 3q93_B |
| 1hx6_C | 1oot_A | 1vd6_A | 2bf6_A | 2hba_A | 2qrw_I | 2z1c_B | 3edf_A | 3jzy_A | 3qan_C |
| 1hxh_D | 1oqj_A | 1ve1_A | 2bfw_A | 2hbv_A | 2qsa_A | 2z2f_A | 3edv_A | 3k01_A | 3qat_B |
| 1hxr_B | 1oqv_C | 1vef_B | 2bgs_A | 2hc8_A | 2qsk_A | 2z38_A | 3ee4_A | 3k1h_A | 3qby_A |
| 1hz4_A | 1orn_A | 1vfl_A | 2bh8_B | 2hc9_A | 2qsq_B | 2z3g_D | 3eeh_A | 3k26_A | 3qc7_A |
| 1hz6_C | 1orr_A | 1vfr_B | 2bii_B | 2hd9_A | 2qt7_B | 2z3v_A | 3ees_A | 3k2w_E | 3qds_B |
| 1hzj_A | 1os6_A | 1vfy_A | 2bjd_B | 2hda_A | 2qub_I | 2z4u_A | 3ef4_A | 3k31_A | 3qe1_A |
| 1hzo_A | 1osy_B | 1vg8_C | 2bjf_A | 2he0_A | 2qud_A | 2z66_B | 3ef6_A | 3k3c_D | 3qgz_A |
| 1hzt_A | 1oth_A | 1vh5_A | 2bji_A | 2he2_A | 2qul_C | 2z6n_A | 3efy_A | 3k3k_A | 3qh4_A |
| 1i0l_A | 1ou8_B | 1vht_B | 2bjq_A | 2he4_A | 2quo_A | 2z6o_A | 3eg4_A | 3k3v_A | 3qhz_M |
| 1i0r_A | 1ouw_C | 1vhw_A | 2bk9_A | 2hek_B | 2quy_H | 2z6r_A | 3egg_D | 3k62_A | 3qk8_C |
| 1i0v_A | 1ov3_A | 1vi6_B | 2bka_A | 2heu_B | 2qv5_A | 2z6w_A | 3ego_B | 3k6f_A | 3qki_B |
| 1i1n_A | 1ow3_A | 1vim_A | 2bkf_A | 2hew_F | 2qvb_A | 2z72_A | 3egw_C | 3k6i_A | 3qmd_A |
| 1i24_A | 1owf_A | 1vj2_A | 2bkl_B | 2hf9_A | 2qvo_A | 2z79_B | 3ehg_A | 3k6v_A | 3qp4_A |
| 1i27_A | 1ox0_A | 1vjk_A | 2bkm_B | 2hfn_H | 2qvu_B | 2z7f_E | 3ehw_B | 3k6y_A | 3qqi_B |
| 1i2t_A | 1oxj_A | 1vjw_A | 2bko_A | 2hgx_B | 2qwc_A | 2z7f_I | 3ei9_B | 3k7f_B | 3qry_A |
| 1i4u_A | 1oxs_C | 1vk5_A | 2bkr_A | 2hhv_A | 2qwl_A | 2z84_A | 3eif_A | 3k7i_B | 3qu1_B |
| 1i6m_A | 1oyg_A | 1vkc_A | 2bkx_A | 2hin_B | 2qwo_B | 2z8f_A | 3ej9_B | 3k7p_B | 3qug_A |
| 1i77_A | 1oz9_A | 1vke_F | 2bky_B | 2hjv_A | 2qx8_B | 2z8l_A | 3ej9_C | 3k89_A | 3qxc_A |
| 1i7h_A | 1ozn_A | 1vki_A | 2bky_Y | 2hke_B | 2qxi_A | 2z8q_A | 3eja_A | 3k8d_A | 3qy1_B |
| 1i7k_A | 1ozw_B | 1vkk_A | 2bl0_A | 2hl7_A | 2qy1_B | 2z8u_B | 3ejf_A | 3k8u_A | 3qyj_A |
| 1i8a_A | 1p0f_B | 1vl1_A | 2bl0_B | 2hlc_A | 2qy9_A | 2z8x_A | 3ejg_A | 3k8w_A | 3qzb_A |
| 1i8f_F | 1p1j_A | 1vl7_A | 2bl8_B | 2hls_A | 2qzt_B | 2z9v_B | 3eju_A | 3k9o_A | 3r0p_B |
| 1i8k_B | 1p1m_A | 1vlc_A | 2blf_A | 2hlv_A | 2r0b_A | 2za0_A | 3eki_A | 3k9w_A | 3r1i_A |
| 1i8o_A | 1p1x_B | 1vlj_A | 2blf_B | 2hmq_D | 2r0h_C | 2zbo_A | 3elw_A | 3kbf_A | 3r1w_C |
| 1i9c_A | 1p28_B | 1vm9_A | 2bme_B | 2hor_A | 2r16_A | 2zbt_B | 3elx_A | 3kcc_A | 3r3r_A |
| 1iap_A | 1p3c_A | 1vmb_A | 2bmo_A | 2hos_B | 2r1j_R | 2zc8_A | 3em1_A | 3kcg_H | 3r3s_C |
| 1iby_B | 1p6o_B | 1vmf_C | 2bmo_B | 2hq6_A | 2r2y_A | 2zd1_A | 3emi_A | 3kci_A | 3r6f_A |
| 1idp_A | 1p71_A | 1vmj_A | 2bnm_B | 2hqh_C | 2r31_A | 2zdh_A | 3emw_A | 3kcp_A | 3sil_A |
| 1ig3_A | 1p99_A | 1vp2_A | 2bo1_A | 2hqs_H | 2r37_A | 2zdo_B | 3enb_A | 3kda_A | 4ubp_A |
| 1igq_A | 1pa2_A | 1vp6_C | 2bo4_F | 2hqy_A | 2r5o_B | 2zdr_A | 3enk_B | 3ke4_A | 4ubp_B |
| 1ihj_B | 1pam_B | 1vph_E | 2bo9_C | 2hra_A | 2r6j_B | 2zex_A | 3enu_A | 3kef_B | 4vub_A |
| 1iib_B | 1pcf_C | 1vps_A | 2bo9_D | 2hrv_B | 2r75_1 | 2zez_B | 3eoi_A | 3keo_B | 5pal_A |
| 1ijb_A | 1pdo_A | 1vq3_B | 2boo_A | 2hsa_A | 2r8e_E | 2zfc_B | 3epr_A | 3kfa_A | 6cel_A |
| 1ijt_A | 1pe9_B | 1vqe_A | 2bpd_B | 2ht9_B | 2r8o_A | 2zfd_A | 3eqn_B | 3kff_A | 6rxn_A |
| 1ijx_C | 1pfb_A | 1vsr_A | 2bpq_A | 2hta_A | 2r8q_A | 2zfz_D | 3er6_A | 3kg0_C | 7fd1_A |
| 1ijy_B | 1pgv_A | 1vyf_A | 2bqx_A | 2hu9_A | 2r99_A | 2zgq_A | 3era_B | 3kgr_A | 7rsa_A |
| 1ikt_A | 1pj5_A | 1vyo_A | 2br9_A | 2hur_B | 2r9f_A | 2zhj_A | 3erj_A | 3kgz_B | 8abp_A |
| 1io0_A | 1pk3_B | 1vzi_B | 2bsj_A | 2hv8_A | 2ra3_B | 2zhn_A | 3erx_B | 3kh7_A | |
| 1iom_A | 1pkh_A | 1vzy_B | 2bt6_A | 2hv8_E | 2ra4_A | 2zhz_C | 3esg_B | 3kij_C | |
| 1ioo_B | 1pl3_A | 1w0d_A | 2bt9_A | 2hvm_A | 2ra6_B | 2zib_A | 3esl_B | 3kki_A | |
| 1iq6_B | 1pl8_D | 1w0n_A | 2buu_A | 2hvw_C | 2rbk_A | 2zjd_C | 3eu9_C | 3kkq_A | |